\documentclass[aps,twocolumn,floatfix]{revtex4-2}
\usepackage{bm}
\usepackage{dcolumn}
\usepackage{graphicx}
\usepackage{amsmath}
\usepackage{amssymb} 
\usepackage{amsfonts}
\usepackage{float}
\usepackage{xcolor}
\usepackage[capitalise]{cleveref}
\usepackage{multirow}  
\usepackage{makecell} 

\begin{document}
\title{Gapped out-of-phase plasmon modes in alternating-twist multilayer graphene} 
\author{Taehun Kim}
\author{Hongki Min}
\email{hmin@snu.ac.kr}
\affiliation{Department of Physics and Astronomy, Seoul National University, Seoul 08826, Korea}
\date{\today}

\begin{abstract}
We theoretically investigate the plasmon modes of alternating-twist multilayer graphene. In multilayer systems, interlayer coupling gives rise to distinctive plasmon modes, but calculations in moir\'e systems remain challenging due to their complex tunneling structures. Using the Kac-Murdock-Szeg\H{o} Toeplitz formalism, we derive that the in-phase mode exhibits the conventional $\sqrt{q}$ behavior, while the out-of-phase modes acquire plasmon gaps determined by specific interband transitions between Dirac cones with different velocities in the long-wavelength limit. We demonstrate that these out-of-phase modes remain undamped in the weak Coulomb-interaction limit when the twist angle exceeds a critical value ($\theta \gtrsim 2.75^\circ$ for the alternating-twist trilayer case), regardless of the carrier density as long as the low-energy effective Dirac Hamiltonian remains valid. Furthermore, we consider the effect of a perpendicular electric field, and demonstrate how plasmon modes can be tuned by a gate voltage.
\end{abstract}

\maketitle

\section{\label{sec: Introduction}Introduction}

Twisted systems have been the focus of intense interest following the observation of superconductivity and correlated insulating phases in magic-angle twisted bilayer graphene (TBG) \cite{Cao2018a, Cao2018b, Yankowitz2019, Lu2019}. TBG consists of two graphene sheets separated by an interlayer distance of $d \approx 3$~\r{A} and rotated by a twist angle $\theta$ \cite{Lopes2007, Shallcross2008, Mele2010, Li2010, Shallcross2010, Bistritzer2010, Lopes2012}. The resulting spatially varying interlayer coupling gives rise to a long-period moir\'e superlattice, which in turn produces interesting electronic properties that are strongly dependent on the twist angle. In particular, nearly flat bands emerge at the so-called magic angle $\theta \approx 1.1^\circ$ \cite{Trambly2010, Suarez2010, Bistritzer2011, Tarnopolsky2019}, providing an ideal platform to study correlated physics, where the suppression of kinetic energy due to band flattening enhances the role of electron-electron interactions.

In recent years, plasmon modes, which are collective charge oscillations, in TBG have attracted great attention, both theoretically and experimentally \cite{Stauber2013, Stauber2016, Lewandowski2019, Novelli2020, Stauber2018, Hesp2021, Huang2022}. Experimentally, near-field microscopy and Fourier transform infrared spectroscopy have demonstrated the existence of new plasmon modes in TBG, including chiral and slow plasmons \cite{Huang2022, Hesp2021}. Furthermore, owing to its moir\'e band structure and the layer degree of freedom, TBG is theoretically predicted to host an angle-tunable acoustic plasmon \cite{Cavicchi2024}. Unlike other unconventional plasmon modes that originate from the peculiarity of the Dirac cone itself \cite{Gangadharaiah2008, Ross-Harvey2025}, these modes arise from the moir\'e structure of TBG, thereby unveiling new electromagnetic dynamics and exemplifying TBG as a unique quantum-optical platform.

\begin{figure}[htb]
\includegraphics[width=\linewidth]{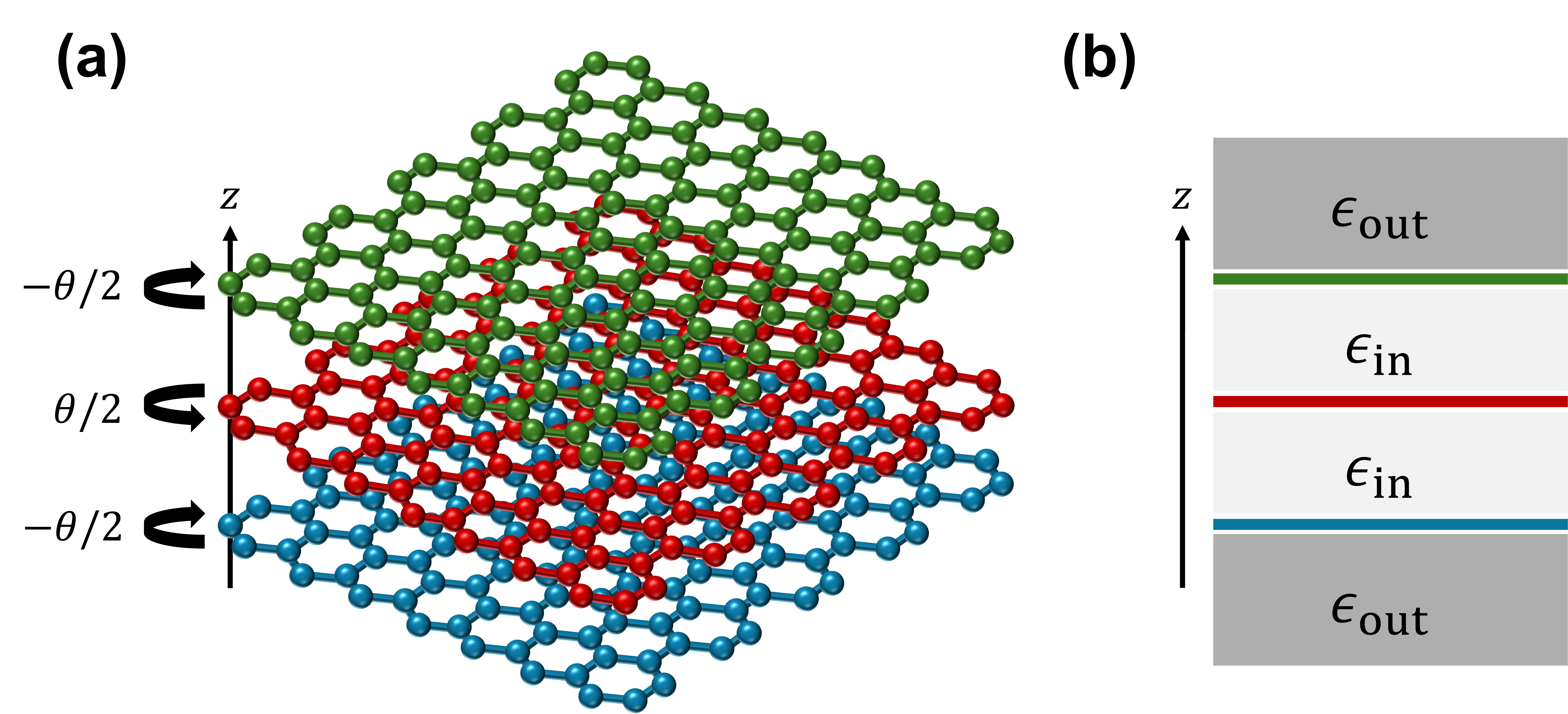}
\caption{
Schematic illustrations of (a) the ATMG with $N=3$ layers and (b) a front view of the ATMG embedded in a dielectric environment with dielectric constants $\epsilon_{\rm in}$ (inside) and $\epsilon_{\rm out}$ (outside). Here, $z$ represents the out-of-plane direction.
} 
\label{fig: fig1}
\end{figure}

In this regard, beyond TBG, collective excitations in alternating-twist multilayer graphene (ATMG) present an intriguing problem. Theoretically, in decoupled $N$-layer systems, there is one in-phase mode with the conventional dispersion ($\omega\propto \sqrt{q}$) and $N-1$ out-of-phase modes with linear dispersions ($\omega\propto q$) \cite{Zhu2013, Kim2020, VanMen2021, Kim2025}. When isotropic interlayer tunneling is present, the out-of-phase modes develop plasmon gaps whereas the in-phase mode remains unaffected \cite{DasSarma1998, Kim2025}. However, the effects of interlayer tunneling in moir\'e systems on these out-of-phase modes remain an open question due to their complex structures.

In this paper, we theoretically analyze how interlayer interactions in ATMG affect plasmon modes. Using the Coulomb-eigenvector basis, which can be obtained exactly from Kac-Murdock-Szeg\H{o} (KMS) Toeplitz matrices \cite{Kac1953, Trench2001, Bogoya2016, Fikioris2018, Narayan2021, Kim2025}, we find that the in-phase mode exhibits the conventional $\sqrt{q}$ dispersion, whereas the out-of-phase modes develop plasmon gaps governed by specific interband transitions between Dirac cones with different velocities. We also investigate the effect of a perpendicular electric field, which splits the Dirac nodes and yields gate-tunable plasmon gaps.

This paper is organized as follows. In Sec. II, we introduce the random phase approximation (RPA) for multilayer systems and the KMS Toeplitz formalism. In Sec. III, we approximately derive the plasmon modes from the low-energy effective Hamiltonian of ATMG in the long-wavelength limit and compare our theoretical results with full numerical calculations. Sec. IV addresses the role of an applied perpendicular electric field in ATMG. Finally, in Sec. V, we conclude with discussions on the effect of the number of layers, the dependence on the twist angle, and the validity of our model with respect to the electric field and carrier density.

\section{\label{sec: RPA theory applied to multilayer systems}RPA theory applied to multilayer systems}

We consider a single-particle Hamiltonian $H(\bm{k})$ for an $N$-layer system with interlayer separation $d$, assuming that the material is embedded between two dielectric media of dielectric constant $\epsilon_{\rm out}$, while the space between the layers is filled with another dielectric medium of dielectric constant $\epsilon_{\rm in}$ (see Fig.~\ref{fig: fig1}). The corresponding eigenstates are denoted by $|\bm{k}, \lambda\rangle$ with energy dispersions $\varepsilon_{\bm{k}, \lambda}$ and the band index $\lambda$. Throughout this paper, we set $\hbar = 1$.

To describe the collective modes in multilayer systems, we consider the imaginary-time response function in the layer basis as \cite{Pines1966, Mahan2000, Giuliani2005, Kamatani2022}
\begin{equation}
\begin{split}
    \chi_{ij}^{(0)}(\bm{q}, i\Omega_m) = & \frac{1}{\beta} \sum_{i\omega_n} \frac{g}{V} \sum_{\bm{k}} \text{tr}[G(\bm{k}, i\omega_n)P_{i} \\
    & \times  G(\bm{k}+\bm{q}, i\omega_n+i\Omega_m)P_{j}].
\end{split}
\end{equation}
Here, $g$ is the spin-valley degeneracy, $i, j$ are the layer indices, $P_{i}$ is the projection operator on the $i$th layer, $\omega_n = (2n+1) \pi / \beta$ and $\Omega_m = 2 m \pi / \beta$ are the fermionic and bosonic Matsubara frequencies with $\beta = 1/k_{\rm{B}} T$, where $k_{\rm{ B}}$ denotes the Boltzmann constant and $T$ is the temperature. The Green's function is defined as
\begin{equation}
    G(\bm{k}, i\omega) = \frac{1}{i\omega - H(\bm{k})} = \sum_\lambda \frac{|\bm{k}, \lambda \rangle\langle\bm{k}, \lambda|}{i\omega - \varepsilon_{\bm{k}, \lambda}}.
\end{equation}
Performing the Matsubara frequency summation, and taking the analytic continuation $i\Omega_m \rightarrow \omega + i\eta$ (with $\eta$ being a positive infinitesimal), we obtain the noninteracting density-density response function as 
\begin{equation} \label{eq: noninteracting_response-layer_basis}
\begin{split}
    \chi_{ij}^{(0)}(\bm{q}, \omega) =& \frac{g}{V}\sum_{\bm{k}, \lambda, \lambda'}\frac{f_{\bm{k}, \lambda}-f_{\bm{k+\bm{q}}, \lambda'}}{\omega + \varepsilon_{\bm{k}, \lambda} - \varepsilon_{\bm{k+\bm{q}}, \lambda'}+i\eta}\\
    &\times F_{ij}^{\lambda\lambda'}(\bm{k}, \bm{k+\bm{q}}),
\end{split}
\end{equation}
where $F_{ij }^{\lambda\lambda'}(\bm{k},\bm{k}')=\langle\bm{k}, \lambda|P_i|\bm{k}', \lambda'\rangle \langle \bm{k}', \lambda'|P_j|\bm{k}, \lambda\rangle$ is the overlap factor and $f_{\bm{k}, \lambda}= [e^{\beta(\varepsilon_{\bm{k}, \lambda}-\mu)}+1]^{-1}$ is the Fermi-Dirac distribution at the chemical potential $\mu$. This density-density response function is fundamentally linked to the layer basis current-current response function via the continuity equation \cite{Dionisios2024, Vasilevskiy2026}. We set $T = 0$ K in this paper, so that the Fermi distribution simplifies to a step function $f_{\bm{k}, \lambda} = \Theta(\mu - \varepsilon_{\bm{k}, \lambda} )$.

The matrix elements of the RPA dielectric function in the layer basis can be calculated as
\begin{equation}
    \begin{split}
        \epsilon_{ij}(\bm{q}, \omega) = \delta_{ij} - \sum_{k}V_{ik}(q)\chi_{kj}^{(0)}(\bm{q}, \omega).
    \end{split}
\end{equation}
\noindent Here, the elements of the Coulomb matrix $V_{ij}(q)$ represent the Coulomb interactions between the $i$th and $j$th layers with $q = |\bm{q}|$. The bare Coulomb interactions, influenced by the dielectric environment, can be obtained from a straightforward electrostatics calculation \cite{Profumo2010}. We find that, in the long-wavelength limit, the Coulomb matrix eigenvalues $V_{\alpha}(q)$ and the corresponding eigenvectors $\bm{u}_\alpha(q) = \frac{1}{N_\alpha} \big[u_{\alpha}^{(1)}(q), u_{\alpha}^{(2)}(q), \ldots, u_{\alpha}^{(N)}(q)\big]^T$ with the normalization constant $N_\alpha$ ($\alpha = 1, 2, \ldots, N$) take the form of those in a special case of a Toeplitz matrix known as the KMS matrix \cite{Kac1953, Trench2001, Bogoya2016, Fikioris2018, Narayan2021, Kim2025}. By solving the KMS matrix, the Coulomb eigenvalues and eigenvectors in the long-wavelength limit are obtained as
\begin{subequations}\label{eq:rho->1 limit}
    \begin{flalign}
        V_\alpha(q\rightarrow 0) &= \frac{2\pi e^2}{q}\times\begin{cases}
        \frac{N}{\epsilon_{\rm{out}}},&  \alpha=1, \\ 
        \frac{1}{\epsilon_{\rm{in}}}\frac{qd}{1-\cos(\frac{\alpha-1}{N}\pi)}, & \alpha\neq1,
        \end{cases}\\
        u_{\alpha}^{(k)}(q \rightarrow 0) &= \cos\left[\frac{(2k-1)(\alpha-1)}{2N}\pi\right].
    \end{flalign}
\end{subequations}
\noindent  We leave the details in Appendix~\ref{sec: Multilayer Coulomb interaction}. Note that the Coulomb eigenvalue of the in-phase mode ($\alpha = 1$) is governed primarily by the dielectric constant $\epsilon_{\rm{out}}$, while those of the out-of-phase modes ($\alpha \neq 1$) are governed by $\epsilon_{\rm{in}}$.

In the Coulomb eigenvector basis, the dielectric function can be expressed as \cite{Kim2025}
\begin{equation}
    \begin{split}
        \epsilon_{\alpha\beta}(\bm{q}, \omega) = \delta_{\alpha\beta} - V_\alpha(q)\chi^{(0)}_{\alpha\beta}(\bm{q}, \omega).
    \end{split}
    \label{dielectric_Coulomb}
\end{equation}
\noindent Here, $\chi^{(0)}_{\alpha\beta}(\bm{q}, \omega)$ denotes the noninteracting density-density response function in the Coulomb eigenvector basis. It can be obtained from Eq.~(\ref{eq: noninteracting_response-layer_basis}) by replacing the overlap factor $F_{ij}^{\lambda\lambda'}(\bm{k}, \bm{k'})$ with
$F_{\alpha\beta}^{\lambda\lambda'}(\bm{k}, \bm{k'}) = \langle\bm{k}, \lambda|U_{\alpha}|\bm{k'}, \lambda'\rangle \langle\bm{k'}, \lambda'|U_\beta|\bm{k}, \lambda\rangle$, where $U_\alpha = \text{diag}[\bm{u}_{\alpha}(q)\otimes\mathbb{I}]$ with the identity matrix $\mathbb{I}$ of the single layer. Note that the Coulomb eigenvectors alternate in parity: those with odd indices are symmetric with respect to the midpoint, while those with even indices are antisymmetric. Unlike the layer basis, which requires full matrix diagonalization, the Coulomb eigenbasis imposes strong selection rules for interband transitions that make analytical treatment more tractable. In particular, for systems whose bands are either symmetric or antisymmetric, interband transitions between symmetric and antisymmetric bands occur only via antisymmetric Coulomb modes, whereas all other transitions proceed through symmetric modes.

\section{\label{sec: Collective excitations of ATMG}Collective excitations of ATMG}

\subsection{Continuum model for ATMG}

We consider a continuum Hamiltonian for an $N$-layer graphene system in the moir\'e Brillouin zone (mBZ), where the $i$th layer is twisted alternately by an angle $\theta_{i} = (-1)^i \theta / 2$. The Hamiltonian of ATMG is given by \cite{Khalaf2019, Leconte2022, Shin2023}
\begin{eqnarray}\label{eq: Hamiltonian of ATMG}
    H = \begin{pmatrix}
        h^{(1)}(\bm{k}) & T(\bm{r}) & 0 &\cdots \\
        T^{\dagger}(\bm{r}) & h^{(2)}(\bm{k}) & T^{\dagger}(\bm{r}) & \cdots \\
        0 & T(\bm{r}) & h^{(3)}(\bm{k}) & \cdots \\
        \vdots & \vdots & \vdots & \ddots
    \end{pmatrix},
\end{eqnarray}
\noindent where $h^{(i)}(\bm{k}) = v_0 (\bm{k}\cdot\bm{\sigma}_{\theta_{i}})$, with $\bm{\sigma}_{\theta_{i}} = e^{\frac{i}{2}\theta_i \sigma_z} \bm{\sigma} e^{-\frac{i}{2}\theta_i \sigma_z}$, describes the Dirac cones of the twisted layers, and $T(\bm{r})$ denotes the interlayer tunneling. Here, $v_0 = \sqrt{3}|t|a / 2$ is the Fermi velocity of monolayer graphene, with the nearest-neighbor hopping parameter $t = -3.1$~eV and lattice constant $a = 2.46$~\r{A}. The corresponding interlayer hopping can be expressed as
\begin{eqnarray}
    T(\bm{r}) &=& \sum_{j = 0, \pm} e^{i \bm{q}_j\cdot \bm{r}} T_j,
\end{eqnarray}
\noindent where $\bm{q}_{\pm} = 2 k_{\rm{D}} \sin(\theta / 2) (\pm \sqrt{3} / 2, 1/2)^T$ and $\bm{q}_0 = 2k_{\rm{D}} \sin(\theta / 2) (0, -1)^{T}$ represent three equivalent hopping paths between the adjacent layers with $k_{\rm{D}} = 4\pi / 3a$. Taking into account the lattice corrugation arising from out-of-plane relaxation \cite{Jung2014, Uchida2014, vanWijk2015}, the interlayer tunneling matrices are written as
\begin{eqnarray}
    T_0 = \begin{pmatrix}
        w' & w \\ w & w'
    \end{pmatrix}, \quad T_\pm =\begin{pmatrix}
        w' & w e^{\mp i\frac{2\pi}{3}} \\
        w e^{\pm i\frac{2\pi}{3}} & w'
    \end{pmatrix}
\end{eqnarray}
with the intrasublattice (intersublattice) hopping term given by $w' = 93.9$ meV ($w = 120$ meV). For the full numerical calculations, we consider the corrugated lattice structure with unequal hopping strengths ($w \neq w'$), whereas for the analytic calculation we adopt the rigid model with equal hopping terms ($w = w' = 120$ meV) \cite{Chebrolu2019}.

\begin{figure}[htb]
\includegraphics[width=\linewidth]{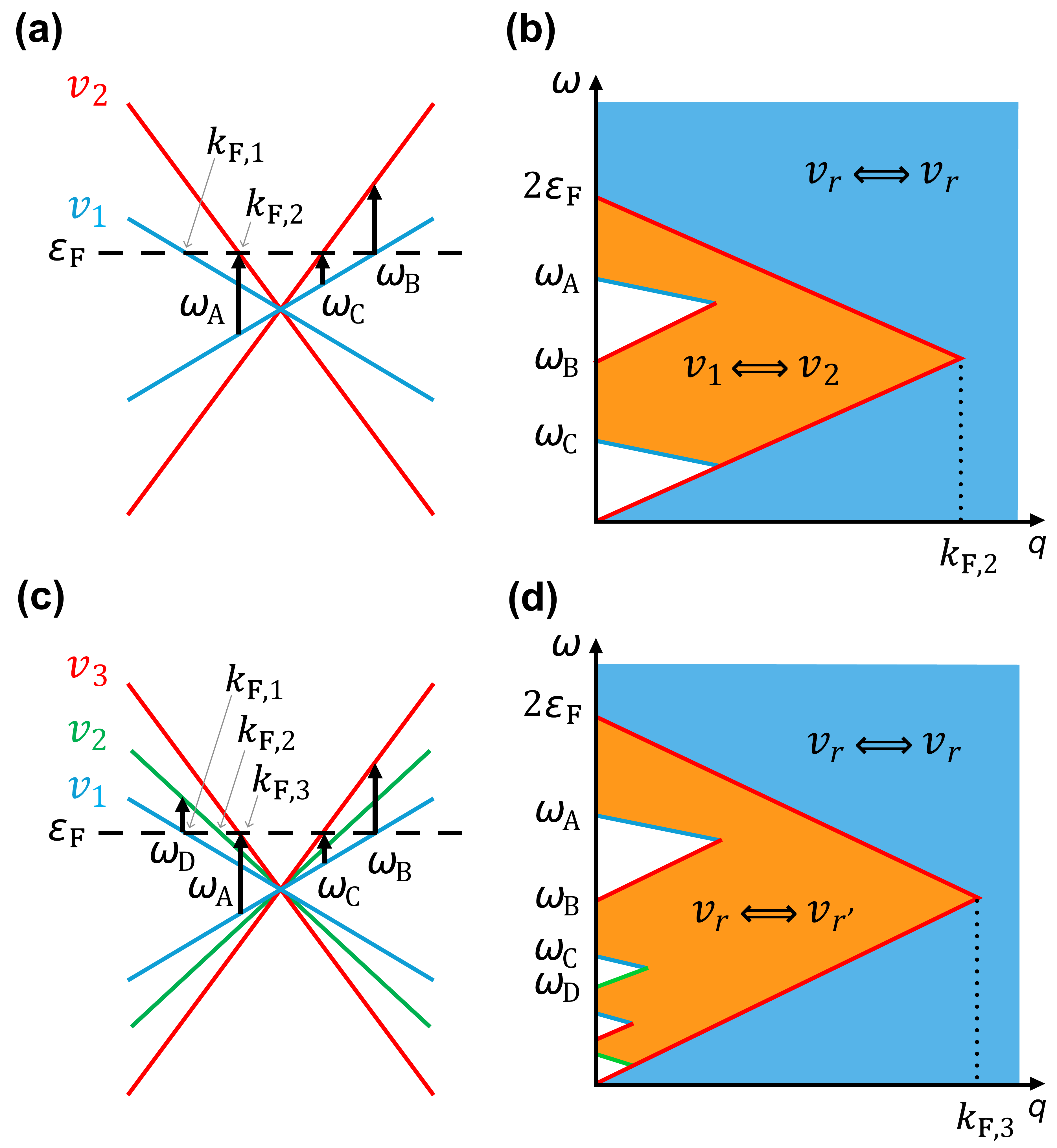}
\caption{Schematic illustrations of the band structures and associated particle-hole continua of ATMG. (a) Linear band dispersions for two Dirac cones with velocities $v_1$ and $v_2$. Dashed black lines denote the Fermi energy $\varepsilon_{\rm{F}}$ and gray arrows represent the Fermi wave vector $k_{\text{F}, r}$. Black arrows depict representative interband transitions at $q = 0$ ($\omega_{\rm{A}}$-$\omega_{{\rm{C}}}$), given by $(v_1 + v_2)k_{\rm{F}, 2}$, $v_{21} k_{\rm{F}, 1}$, and $v_{21} k_{\rm{F}, 2}$, respectively. (b) The resulting particle-hole continuum in the $(q, \omega)$ plane for (a). The orange region indicates transitions between different Dirac cones, whereas the sky-blue regions denote transitions within the same Dirac cone. The boundaries are linear in $v_r q$, with slopes corresponding to the velocities indicated by the colors in (a). (c) Linear band dispersions for three Dirac cones with velocities $v_1$, $v_2$, and $v_3$. Black arrows (labeled $\omega_{\rm{A}}$-$\omega_{\rm{D}}$) depict interband transitions given by $(v_1 + v_3)k_{\rm{F}, 3}$, $v_{31} k_{\rm{F}, 1}$, $v_{31} k_{\rm{F}, 3}$, and $v_{21}k_{\rm{F}, 1}$, respectively. (d) The resulting particle-hole continuum for (c), following the same color scheme as in (b).
} 
\label{fig: fig2}
\end{figure}

The energy levels and corresponding wave functions of ATMG near the two mBZ corners $\bar{K}$ and $\bar{K}'$ can be obtained within the first-shell model, where the moir\'e reciprocal lattice vectors are truncated to the nearest reciprocal vectors and equal sublattice hopping terms ($w = w'$) are assumed. Near $\bar{K}$ or $\bar{K}'$, the low-energy effective eigenfunctions of ATMG can be constructed following the solution of a one-dimensional chain problem as $|\Psi_{r, s}\rangle = \frac{1}{N_{\rm{norm}}} (\Psi_{r, s}^{(1)}, \Psi_{r, s}^{(2)}, \dots, \Psi_{r, s}^{(N)})^T$ with the normalization constant $N_{\rm{norm}}$, where \cite{Khalaf2019, Shin2023}
\begin{eqnarray}\label{eq: egienvector of ATMG}
\Psi_{r, s}^{(k)} = \sin(k\phi_r)\psi_{r, s}.
\end{eqnarray}
\noindent Here, $s=\pm1$ indicates the band index of the Dirac cones, and $\phi_r = \frac{r\pi}{N+1}$ with $r = 1, 2, \dots, n$ for even $N=2n$, or $r = 1, 2, \dots, n+1$ for odd $N = 2n+1$. Detailed expressions for $\psi_{r, s}$ are presented in Appendix~\ref{sec: Wave functions of ATMG}. The corresponding low-energy effective Hamiltonian of ATMG becomes $H_{\rm eff} = v_{r} (\bm{k}\cdot\bm{\sigma})$, where the Fermi velocity $v_{r}$ of the $r$th Dirac cone at $\bar{K}$ or $\bar{K}'$ is given by
\begin{eqnarray}
\frac{v_{r}}{v_0} = \frac{1 - 12 \tilde{\alpha}^2 \cos^2{\phi_r }}{1 + 24\tilde{\alpha}^2\cos^2{\phi_r}}
\end{eqnarray}
\noindent with $\tilde{\alpha} = w/ [2 v_0 k_{\rm{D}}\sin{(\theta/2)}]$ denoting a dimensionless parameter. Note that the velocity of the $r$th Dirac cone is the same in each valley $\bar{K}$ and $\bar{K}'$, and for an odd number of layers, the $(n+1)$th mode near $\bar{K}$ has a velocity $v_{n+1} = v_0$ equal to that of monolayer graphene.

For the full numerical calculations, on the other hand, the Hamiltonian in Eq.~(\ref{eq: Hamiltonian of ATMG}) is expanded in the moir\'e reciprocal lattice without restricting to the first shell. At all twist angles, we retain the reciprocal lattice vectors within a momentum cutoff of $3|\bar{\bm{G}}|$, where $\bar{\bm{G}}$ is a primitive reciprocal lattice vector of the mBZ. This cutoff is sufficient to obtain converged bands for all angles considered in this work. Diagonalizing the Hamiltonian gives the full band structure of ATMG, which reduces to the low-energy effective Dirac cones obtained above near $\bar{K}$ and $\bar{K}'$.

The electronic bands of ATMG are either symmetric or antisymmetric with respect to the middle layer, as they originate from the solutions of a one-dimensional chain problem. Thus, one can use the symmetry properties of the Coulomb eigenbasis, which impose strong selection rules on interband transitions. In the presence of tunneling in ATMG, the plasmon modes derived from the low-energy effective Hamiltonian in the long-wavelength limit take the following form (see Appendix~\ref{sec: Calculation of the noninteracting density-density response function}):
\begin{subequations}\label{eq: in-phase dispersion and out-of-phase mode}
    \begin{flalign}
        \omega_1^2(q\rightarrow 0) &= \frac{2\pi e^2}{\epsilon_{\rm{out}}} \Bigg(\sum_{\nu, r} \frac{n_{\nu, r}}{m_{r}}\bigg)q, \label{eq: in-phase mode dispersion} \\
         \omega_{1<\alpha<\frac{N+2}{2}}^2(q\rightarrow 0) &= \omega_{\rm{gap}, \alpha}^2 + C_\alpha q,
    \end{flalign}
\end{subequations}
\noindent where $n_{\nu, r}$ is the carrier (electron or hole) density of the $r$th Dirac cone at $\nu = \bar{K}$ or $\nu = \bar{K}'$, and $m_{r} = k_{\text{F}, r} / v_{r}$ is the effective mass defined using its Fermi wave vector $k_{\text{F}, r}$ with the Fermi energy $\varepsilon_{\rm{F}} = v_rk_{\text{F},r}$. Here,  $\omega_{\rm{gap}, \alpha}=\omega_{\alpha}(q=0)$ is the gap of the out-of-phase mode with $\alpha = 2, \dots, \frac{N}{2}$ for even $N$ or $\alpha = 2, \dots, \frac{N+1}{2}$ for odd $N$. Notably, these out-of-phase modes are governed by $\epsilon_{\rm{in}}$, whereas the in-phase mode $\omega_1$ depends on $\epsilon_{\rm{out}}$. Approximately, the gap of the out-of-phase mode in the weak Coulomb-interaction limit can be expressed as
\begin{eqnarray}\label{eq: approximated form}
    \omega_{\rm{gap}, \alpha} \approx v_{\alpha1}k_{\text{F}, 1}
\end{eqnarray}
\noindent with $v_{\alpha 1} = v_{\alpha} - v_1$. Here, $v_{\alpha 1} k_{\rm{F}, 1}$ corresponds to the transition energy labeled $\omega_{\rm{B}}$ in Fig.~\ref{fig: fig2}. Unlike the case of isotropic tunneling, which yields one in-phase mode and $N-1$ out-of-phase modes, the above expressions do not capture all Coulomb eigenmodes. Within the scope of the low-energy theory, particularly where the Fermi energy crosses the Dirac cones, the plasmon modes ($\alpha \geq (N+2)/2$) are strongly suppressed, similar to those in AB bilayer and ABA trilayer graphene \cite{Borghi2009, Lin2020}.

\begin{table}[t] 
  \caption{Summary of the critical angles above which the out-of-phase plasmon modes are undamped in the weak Coulomb-interaction limit. Schematic figures show the dominant interband transitions for $\omega_\alpha$ (black arrows) and the Fermi energy $\varepsilon_{\rm{F}}$ (dashed lines) at $\bar{K}$ and $\bar{K}'$ in the long-wavelength limit for a given $N$ in ATMG. For $N=3$ and $N=4$, the critical angle for $\omega_2$ is determined by the condition $v_2 = (1 + \sqrt{2})v_1$. For $N=5$ and $N=6$, $\theta_c$ is defined by $v_3 = (1 + \sqrt{2})v_1$ for $\omega_3$, and by $\frac{v_2}{v_1} + \frac{v_1}{v_3} = 2$ for $\omega_2$.
}  \label{table: table1}
  \centering
  \begin{tabular}{|c||c|c||c|}
    \hline \noalign{\vskip 1pt}
    $N$ & near $\bar{K}$ & near $\bar{K}'$ & $\theta_c$ \\ 
    \hline\hline
    \multirow{1}{*}{$3$}    & \parbox[c][2.4cm][c]{0.35\linewidth}{\centering \includegraphics[width=\linewidth]{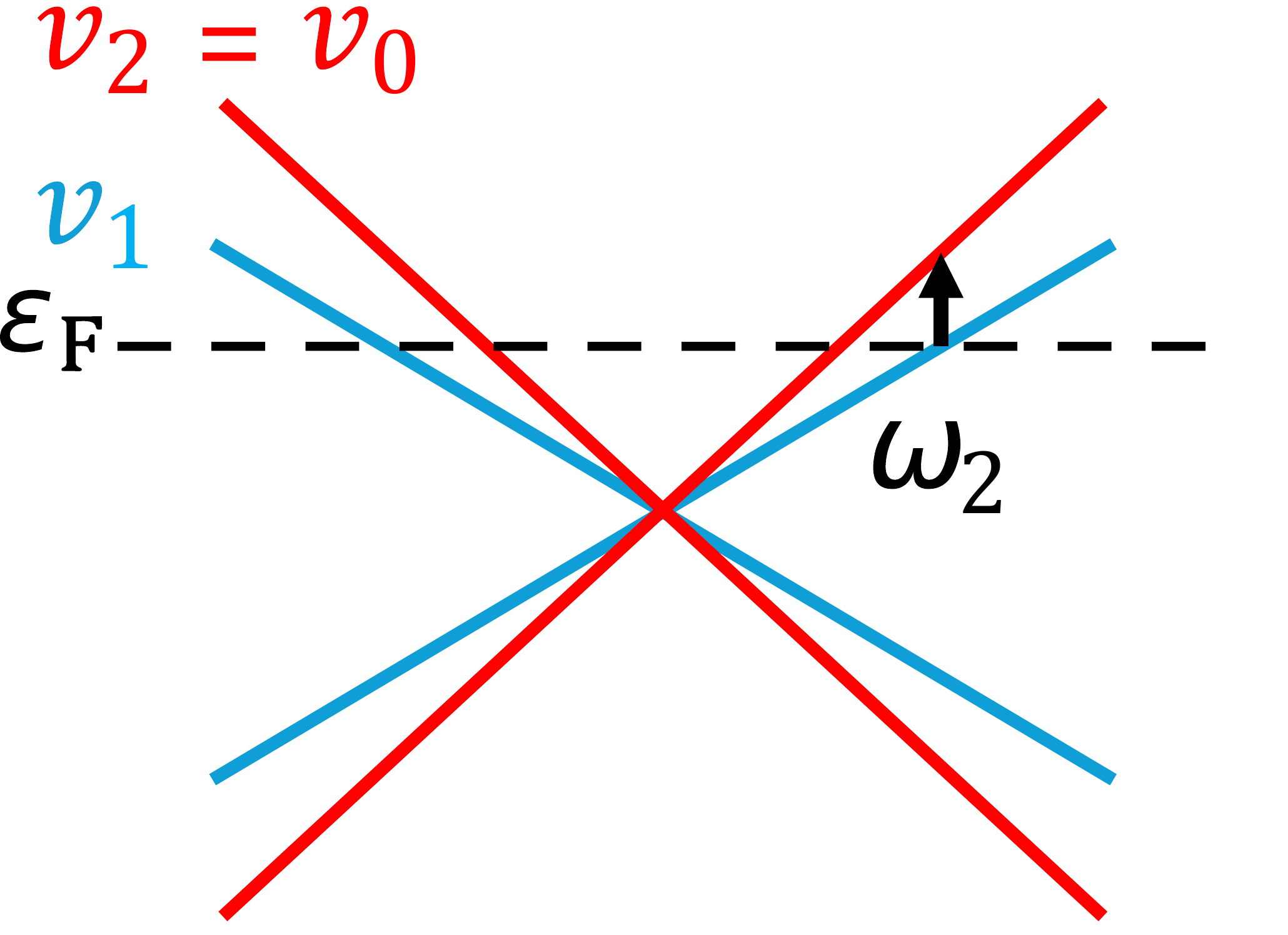}} &  \parbox[c][2.4cm][c]{0.35\linewidth}{\centering \includegraphics[width=\linewidth]{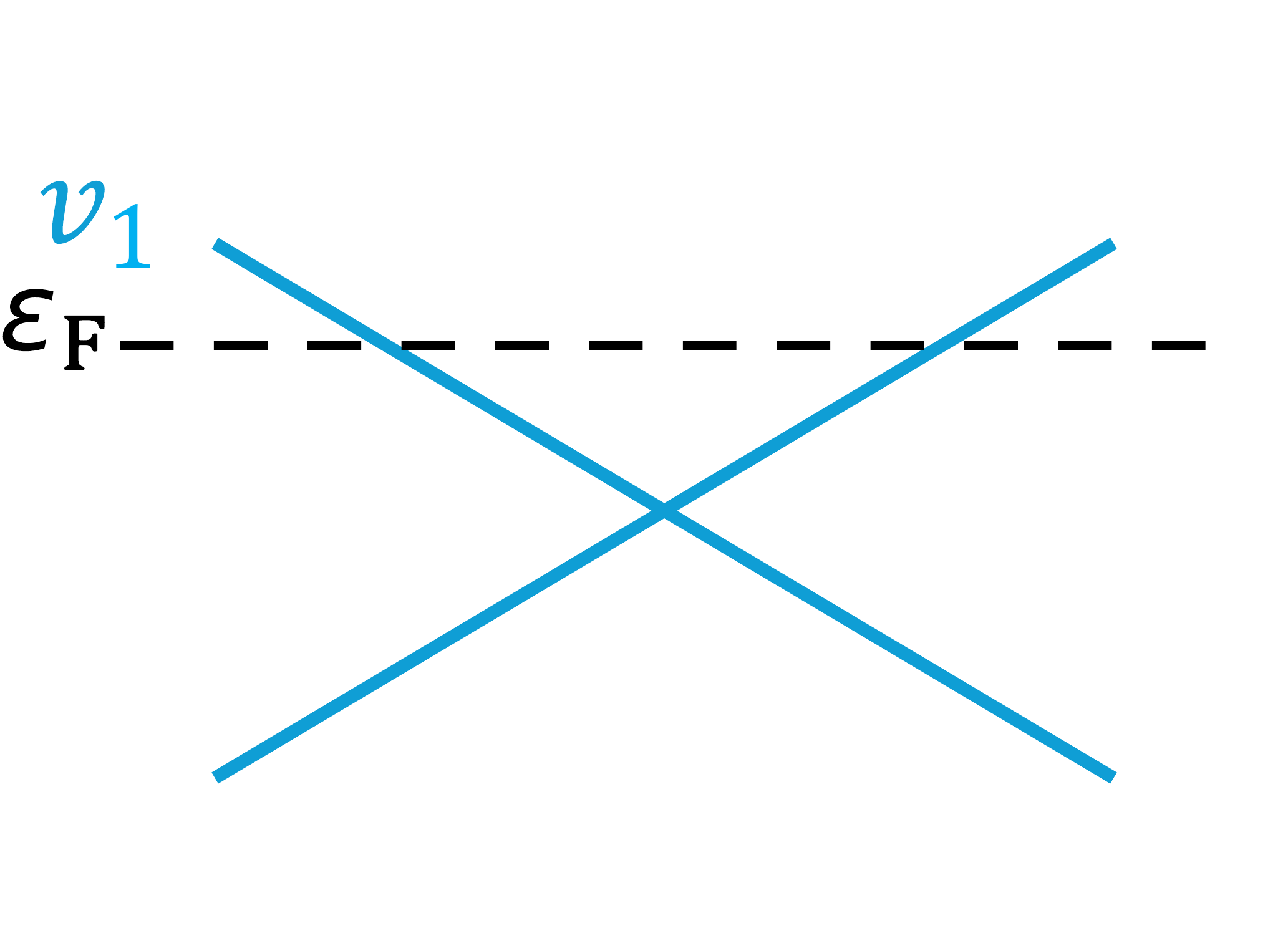}} & $\omega_2$: $2.75^\circ$ \\
    \hline
    \multirow{1}{*}{$4$}     & \parbox[c][2.4cm][c]{0.35\linewidth}{\centering \includegraphics[width=\linewidth]{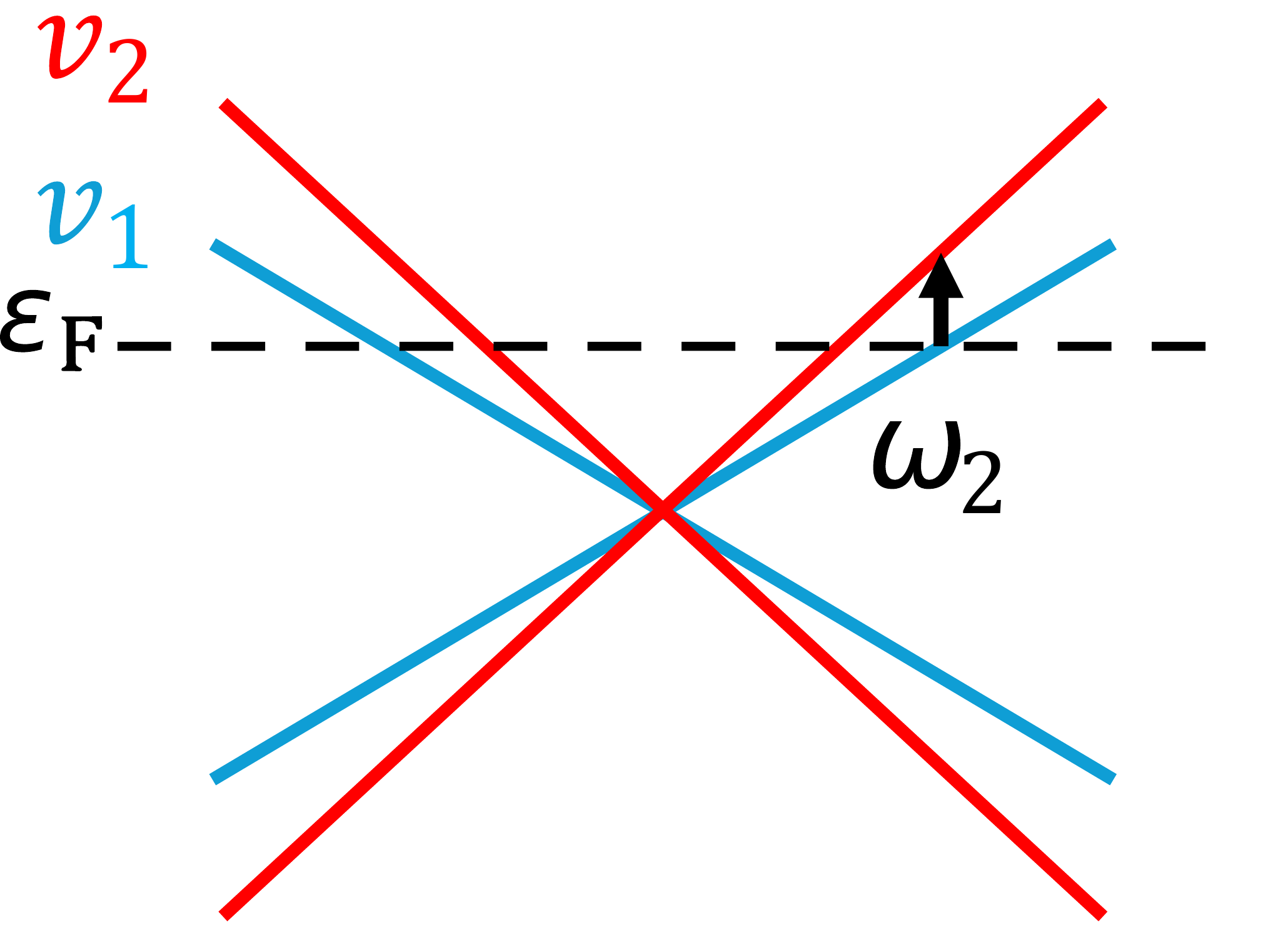}} & \parbox[c][2.4cm][c]{0.35\linewidth}{\centering \includegraphics[width=\linewidth]{t_fig3.pdf}} & $\omega_2$: $2.93^\circ$\\
    \hline
    \multirow{1}{*}{$5$}        & \parbox[c][3.3cm][c]{0.35\linewidth}{\centering \includegraphics[width=\linewidth]{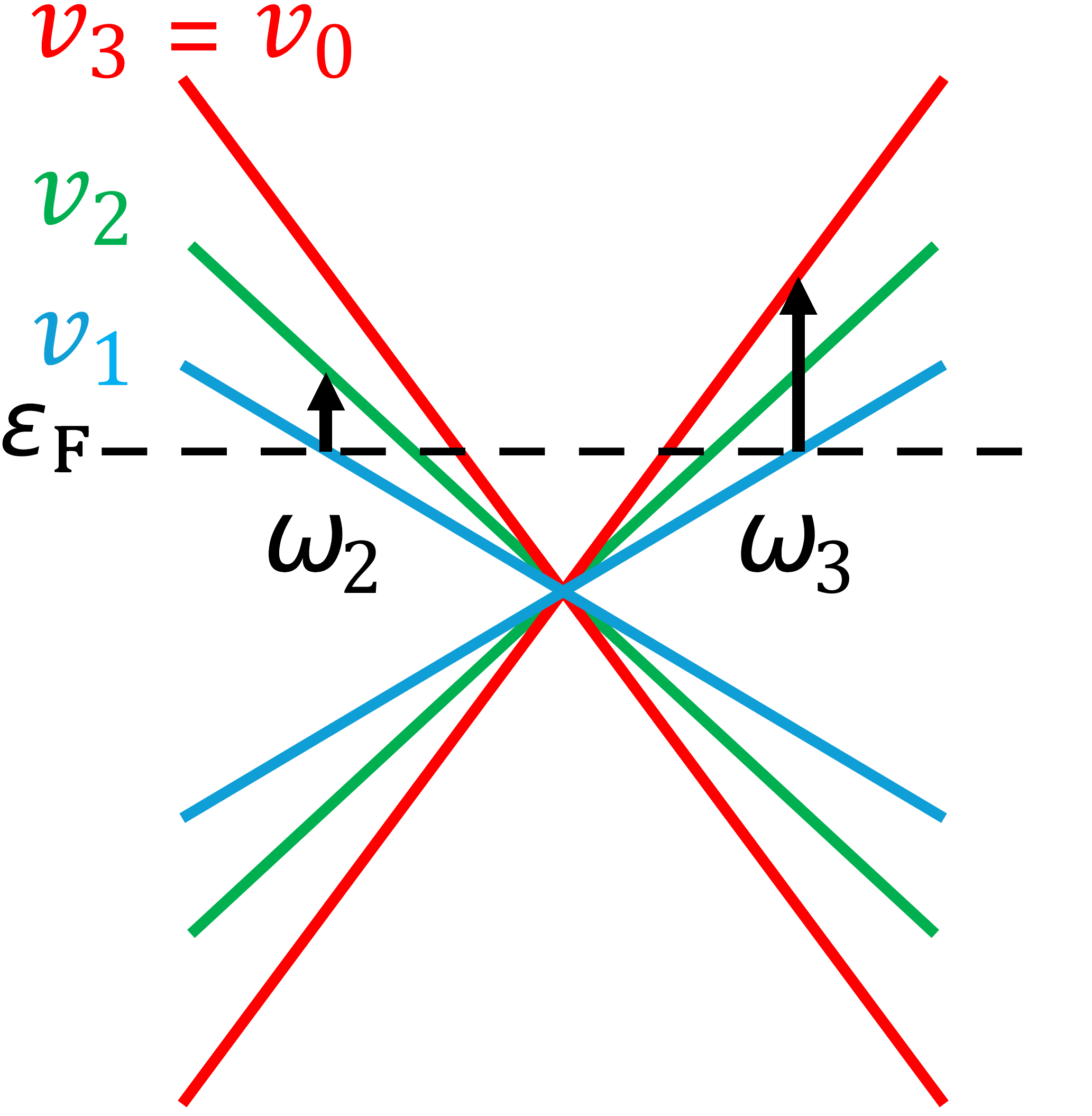}} & \parbox[c][3.3cm][c]{0.35\linewidth}{\centering \includegraphics[width=\linewidth]{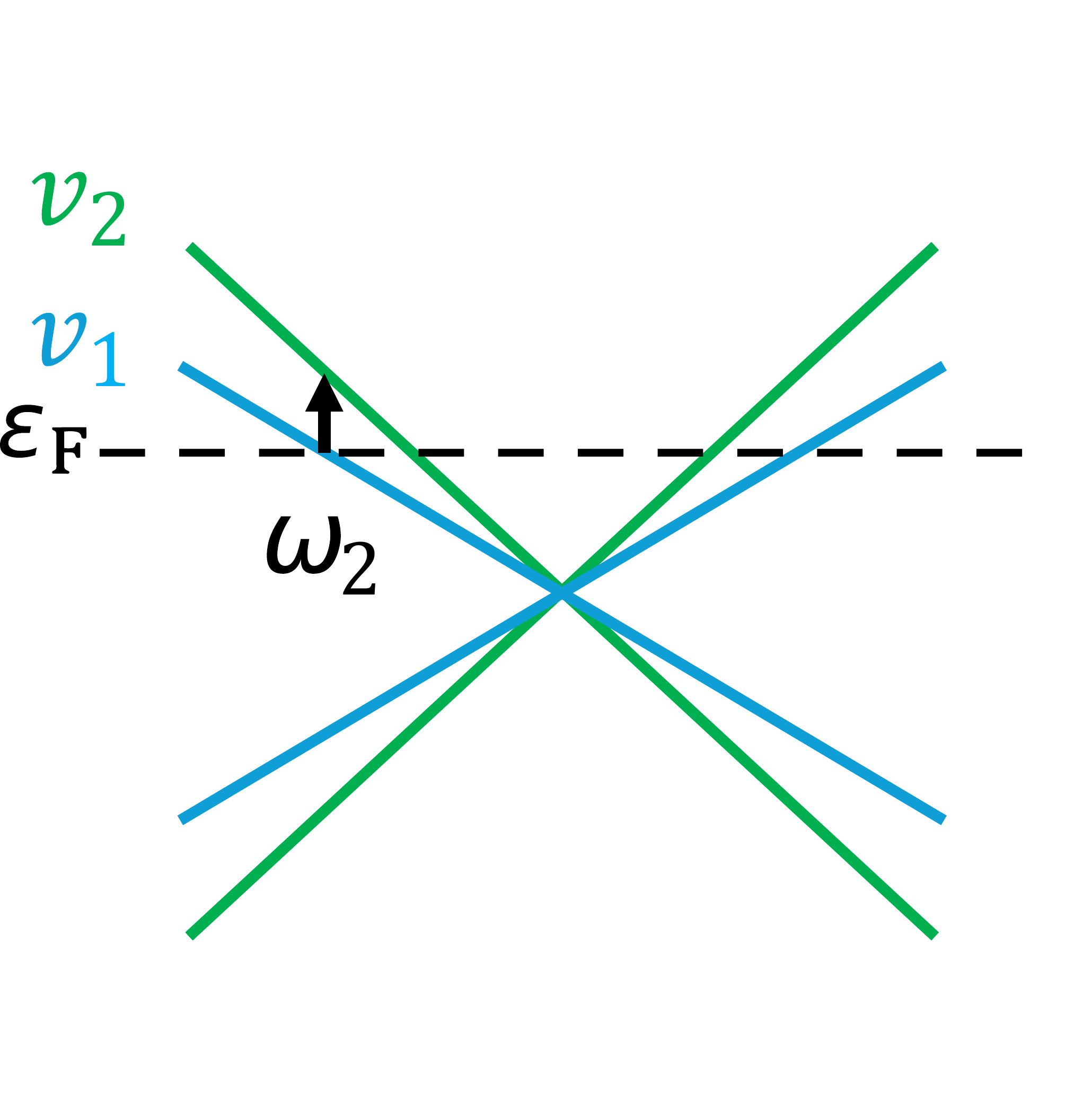}} & \makecell{$\omega_2$: $4.47^\circ$ \\ \\ \\  $\omega_3$: $3.37^\circ$} \\
    \hline
    \multirow{1}{*}{$6$}      & \parbox[c][3.3cm][c]{0.35\linewidth}{\centering \includegraphics[width=\linewidth]{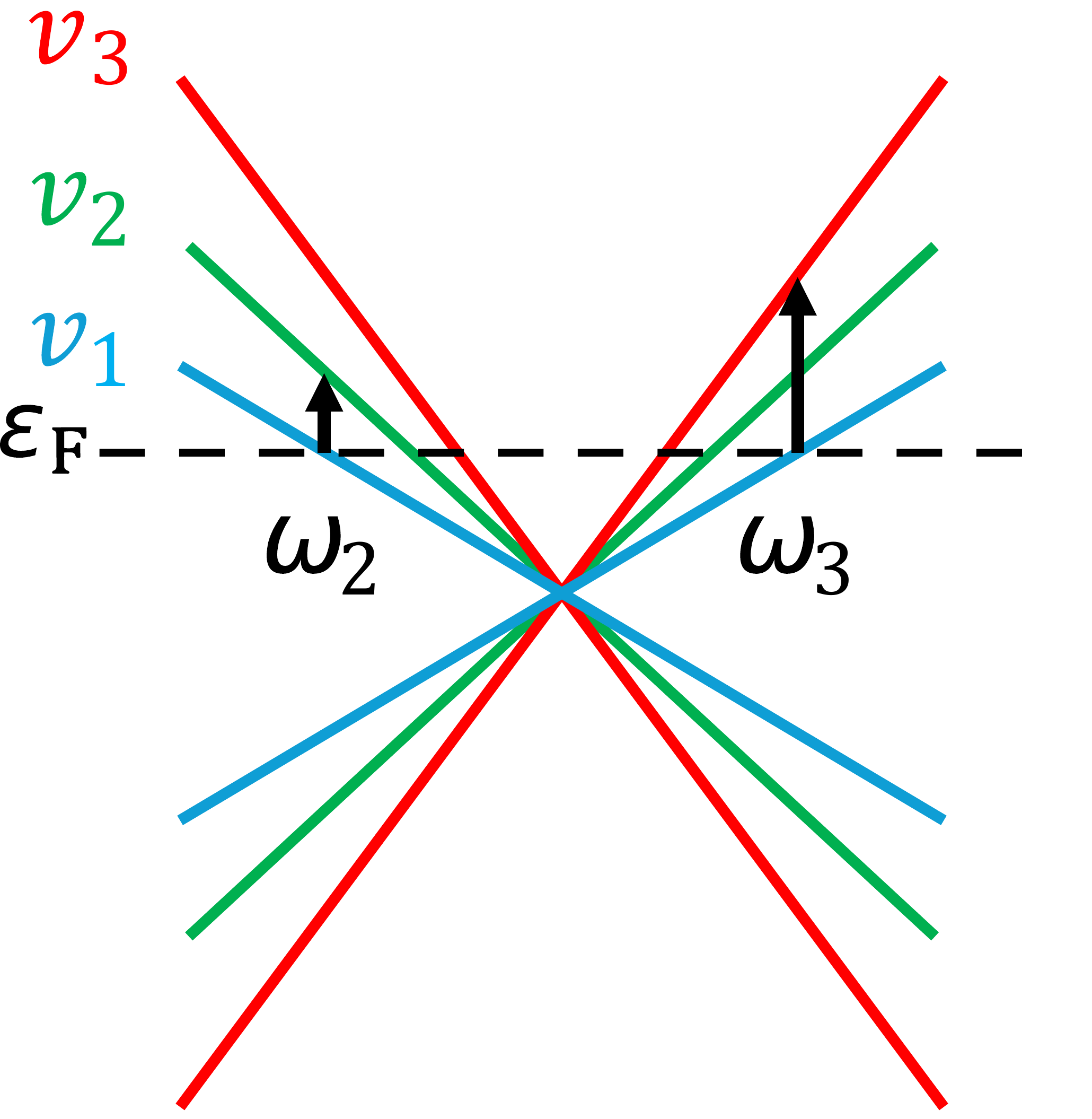}} & \parbox[c][3.3cm][c]{0.35\linewidth}{\centering \includegraphics[width=\linewidth]{t_fig6.pdf}} & \makecell{$\omega_2$: $3.65^\circ$ \\ \\ \\  $\omega_3$: $3.41^\circ$} \\
    \hline
  \end{tabular}
\end{table}

To examine the conditions for Landau damping of out-of-phase plasmon modes, we need to consider the particle-hole continuum of the ATMG. In contrast to monolayer graphene, the presence of distinct velocities $v_r$ allows for interband transitions between different Dirac cones. These transitions form a unique particle-hole continuum in the $(q, \omega)$ plane as shown in Fig.~\ref{fig: fig2}, which constrains the conditions for the formation of long-lived out-of-phase plasmons. Specifically, in the presence of Dirac cones with velocities $v_1$ and $v_2$, the upper boundary of the particle-hole continuum at $q=0$ is given by $\omega_{\rm{A}} = (v_1 + v_2) k_{\rm{F}, 2}$ in Fig.~\ref{fig: fig2}(a), in contrast to the conventional $2\varepsilon_{\rm{F}}$ observed in monolayer graphene. Furthermore, interband transitions between the $v_1$ and $v_2$ Dirac cones give rise to a new particle-hole continuum in the range $\omega_{\rm{C}} < \omega < \omega_{\rm{B}} $ at $q=0$, where $\omega_{\rm{B}} = v_{21}k_{\rm{F}, 1}$ and $\omega_{\rm{C}} = v_{21}k_{\rm{F}, 2}$. 

Based on Eq.~(\ref{eq: approximated form}) and the particle-hole continuum, we can identify the regime where the out-of-phase mode $\omega_\alpha$ ($1<\alpha<\frac{N+2}{2}$) becomes Landau damped. In the weak Coulomb-interaction limit, the out-of-phase mode $\omega_{\alpha}$ with the largest index $\alpha$ remains undamped as long as $\omega_{\rm{gap}, \alpha} < \omega_{\rm{A}}$, where $\omega_{\rm{A}} =  (v_1 + v_\alpha) k_{\rm{F}, \alpha}$. In this case, from $\frac{v_{\alpha}}{v_1} - \frac{v_1}{v_\alpha}< 2$, we obtain $v_{\alpha} < (1 + \sqrt{2})v_1$, which is possible in the high-angle regime and implies that the out-of-phase mode remains undamped once the twist angle exceeds the critical $\theta_{c}$ regardless of the total carrier density $n_{\rm{tot}}$. Analogously, we can determine the critical angles for the remaining out-of-phase modes, which are also independent of $n_{\rm{tot}}$. For instance, the condition for the highest out-of-phase mode to be undamped is always satisfied whenever $\theta >\theta_{c}$ given in Table~\ref{table: table1}, as long as the low-energy effective Dirac Hamiltonian remains valid. Furthermore, for $N=5$ or $N=6$, we find that the out-of-phase mode $\omega_2$ is undamped when $ \omega_{\rm{D}} < \omega_{\rm{C}}$, where $\omega_{\rm{C}} = v_{31} k_{\rm{F}, 3}$ and $\omega_{\rm{D}} = v_{21}k_{\rm{F}, 1}$, leading to the inequality $\frac{v_2}{v_1} + \frac{v_1}{v_3} < 2$. Similarly, we can identify the corresponding critical angle $\theta_c$ above which the mode $\omega_2$ is undamped regardless of $n_{\rm{tot}}$ in the weak Coulomb-interaction limit. Although the actual plasmon gap is shifted by the Coulomb interaction, our findings indicate that the density-independence of the damping conditions is a robust feature of ATMG.

\begin{figure}[b]
\includegraphics[width=1.0\linewidth]{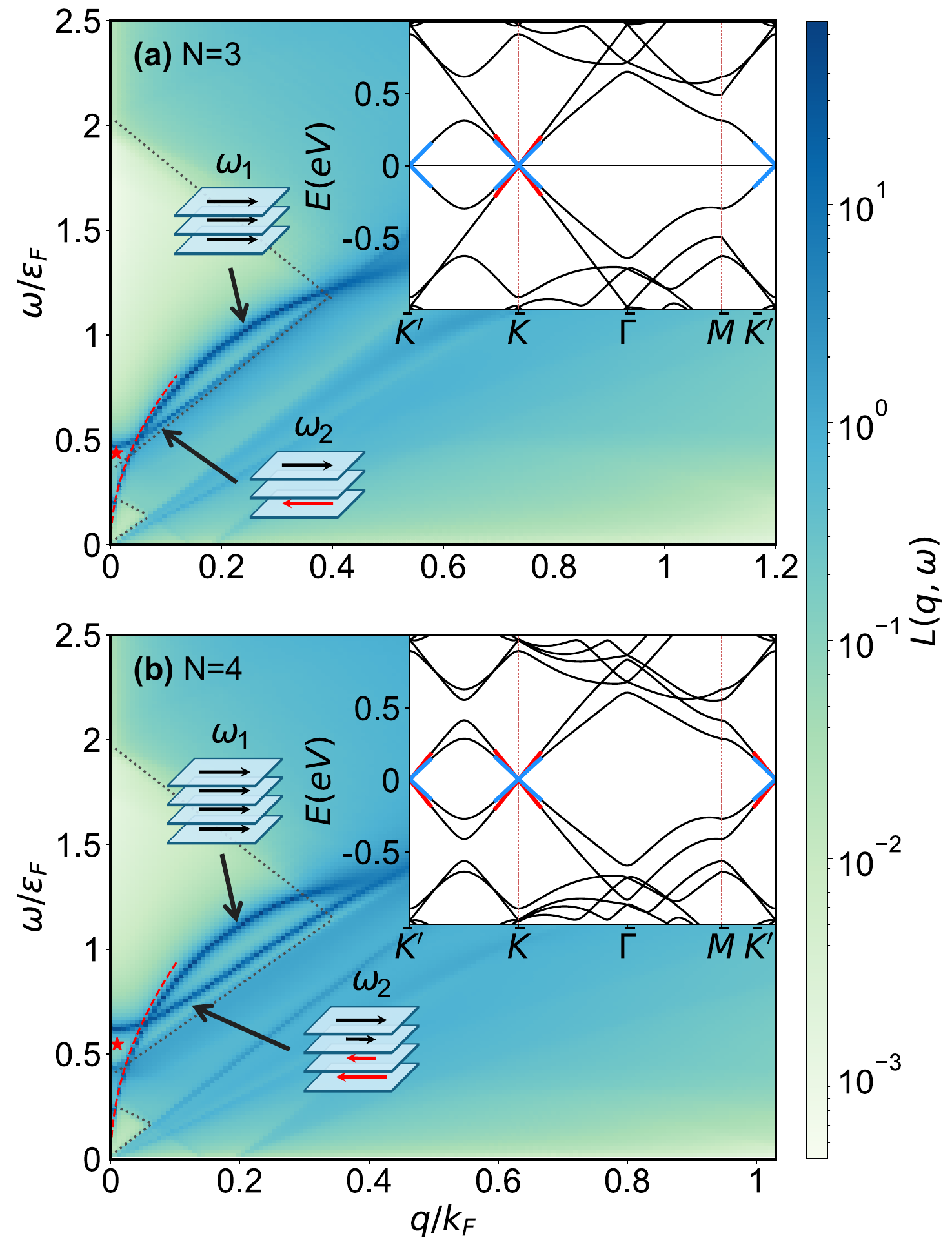}
\caption{Loss function $L(\bm{q}, \omega)$ of unbiased ATMG systems at a twist angle $\theta = 5^{\circ}$ with $\varepsilon_{\text{F}} = 0.2$ eV for (a) $N=3$ ($n_{\rm{tot}} = 1.3 \times 10^{13}$ cm$^{-2}$) and (b) $N=4$  ($n_{\rm{tot}} = 1.8 \times 10^{13}$ cm$^{-2}$). The insets show the band structures of ATMG, where the colored lines highlight the low-energy effective Dirac cones near $\bar{K}$ and $\bar{K}'$. Coulomb eigenvectors of the in-phase and out-of-phase modes are illustrated. The red dashed lines indicate the analytically calculated in-plane plasmon dispersions, and the red stars mark the approximately obtained out-of-phase plasmon gaps using Eq. (15) for $N=3$ and Eq. (16) for $N=4$. The dotted gray lines represent the boundaries of the particle-hole continuum.
}\label{fig: fig3}
\end{figure}

For numerical calculations, we consider the loss function of ATMG defined as \cite{Cavicchi2024, Kim2025}
\begin{equation}\label{eq: loss_eq}
    L(\bm{q}, \omega) = -\text{Im}\{\text{tr}[\epsilon^{-1}(\bm{q}, \omega)]\},
\end{equation}
\noindent which is related to the dynamical structure factor $S(\bm{q}, \omega) \propto L(\bm{q}, \omega)$, representing the spectral weight of electronic excitations. Throughout the main text, we choose $\bm{q}$ along the $\bar{\Gamma}-\bar{M}$ direction, and results along the $\bar{\Gamma}-\bar{K}$ direction are presented in Appendix~\ref{sec: Directional dependence of plasmon dispersion on momentum transfer}. Collective excitations appear as sharp peaks in the loss function, arising from the roots of the equation $\det{[\epsilon(\bm{q}, \omega)]}=0$. All calculations are carried out at a distance $d = 3$ \r{A} and $T=0$ K, using a finite broadening parameter $\eta = 1$ meV and dielectric constants $\epsilon_{\rm{out}} = 5$ and $\epsilon_{\rm{in}} = 1$. In the following, we present the results for the cases $N=3$ and $N=4$, and leave the discussion of the $N=5$ case to Sec. V.

\subsection{$N=3$ case}

Here, we present our calculations for alternating-twist trilayer graphene (AT3G). In the long-wavelength limit, the system exhibits two symmetric oscillation modes [$\bm{u}_{1}\parallel (1, 1, 1)^T$ and $ \bm{u}_3 \parallel (1, -2, 1)^T$] and one antisymmetric oscillation mode [$\bm{u}_2 \parallel (1, 0, -1)^T$]. Figure~\ref{fig: fig3}(a) shows that there are two Dirac cones near $\bar{K}$: one with velocity $v_2=v_{0}$ and an antisymmetric eigenvector, and the other with $v_{1}$ and a symmetric eigenvector, whereas near $\bar{K}'$, there is a single Dirac cone with velocity $v_{1}$ and a symmetric eigenvector.

The numerical results for the plasmon modes are obtained from Eq.~(\ref{eq: loss_eq}) at the twist angle $\theta = 5^\circ$. The in-phase mode ($\bm{u}_1$) arises from intraband transitions within Dirac cones in the mBZ and exhibits a $\sqrt{q}$ dispersion given by Eq.~(\ref{eq: in-phase mode dispersion}). The out-of-phase mode $\omega_2$ arises from interband transitions between the $v_1$ and $v_2 = v_0$ Dirac cones at the $\bar{K}$ point, which involve symmetric and antisymmetric bands. Then, we can approximately obtain the plasmon gap $\omega_{\rm{gap}, 2}$ for the low-energy effective Hamiltonian from the solution of the equation as
\begin{eqnarray}\label{eq: N3omega_gap_2}
    \omega_{\rm{gap}, 2} = v_{21} k_{\rm{F}, 1} + \delta \omega_2^{(3)}(\omega_{\rm{gap}, 2}).
\end{eqnarray}
Here, the additional term $\delta \omega_2^{(3)}$ represents the Coulomb-interaction correction to the out-of-phase mode and is a function of $\omega_{\rm{gap},2}$. See Appendix~\ref{sec: Calculation of the noninteracting density-density response function} for the detailed derivation and the definition of $\delta \omega_2^{(3)}$. Note that the in-phase mode $\omega_{1}$ is symmetric, and thus only symmetric-symmetric or antisymmetric-antisymmetric bands contribute to the oscillation, while the out-of-phase mode $\omega_{2}$ is antisymmetric, permitting only symmetric-antisymmetric bands to contribute to the oscillation. Since they originate from different types of Coulomb modes, they cannot couple to each other.

\subsection{$N=4$ case}
In the case of alternating-twist tetralayer graphene (AT4G), we have two symmetric oscillation modes [$\bm{u}_{1}\parallel (1, 1, 1, 1)^T$ and $ \bm{u}_3 \parallel (1, -1, -1, 1)^T$] and two antisymmetric oscillation modes [$\bm{u}_2 \parallel (1+\sqrt{2}, \sqrt{2}, -\sqrt{2}, -1 - \sqrt{2})^T$ and $\bm{u}_4 \parallel (\sqrt{2}-1, -1, 1, 1 - \sqrt{2})^T$]. At $\bar{K}$ and $\bar{K}'$, there are two Dirac cones with velocities $v_1$ and $v_2$ as illustrated in Fig.~\ref{fig: fig3}(b). Similar to the AT3G case, the eigenvectors of the $v_{1}$ Dirac cones are symmetric, while the eigenvectors of the $v_{2}$ Dirac cones exhibit antisymmetric behavior near both $\bar{K}$ and $\bar{K}'$.

Unlike AT3G, there are two Dirac cones at $\bar{K}'$ in AT4G, and $v_2$ differs from $v_0$. While these features lead to quantitative differences, the in-phase and out-of-phase modes stem from the same underlying mechanism as in AT3G. The in-phase mode ($\bm{u}_1$) arises from intraband transitions within the Dirac cones and exhibits a $\sqrt{q}$ dispersion. The out-of-phase mode $\omega_2$ arises from interband transitions between the $v_{1}$ and $v_{2}$ Dirac cones at both the $\bar{K}$ and $\bar{K}'$ points. The plasmon gap $\omega_{\rm{gap}, 2}$ for the low-energy effective Hamiltonian can be obtained approximately from the solution of the equation as
\begin{eqnarray}\label{eq: N4omega_gap_2}
    \omega_{\rm{gap}, 2} = v_{21} k_{\rm{F}, 1} + \delta \omega^{(4)}_2(\omega_{\rm{gap}, 2}).
\end{eqnarray}
See Appendix~\ref{sec: Calculation of the noninteracting density-density response function} for the definition of $\delta \omega^{(4)}_2$. Just as in the AT3G case, the symmetric mode $\omega_1$ cannot couple to the antisymmetric mode $\omega_2$.

\section{\label{sec: Effect of layer potentials}Effect of layer potentials}

In this section, we consider the plasmon modes of ATMG under a perpendicular electric field. Generally, the effect of an applied field can be effectively modeled through layer-dependent onsite energies by adding a diagonal matrix $H_{\rm{on}} = \text{diag}[\varepsilon_{\rm{on}}^{(1)}, \varepsilon_{\rm{on}}^{(2)}, \dots, \varepsilon_{\rm{on}}^{{(N)}}]$ defined in the layer basis to the ATMG Hamiltonian in Eq.~(\ref{eq: Hamiltonian of ATMG}) \cite{Min2007, Wang2010, Jang2019, Novelli2020, Shin2023, Kolar2023, Cavicchi2024, Kim2025b}. To include the effect of layer potentials induced by an electric field, we use a model parameter $U$ for the layer potentials, where $\varepsilon_{\rm{on}}^{(i)}$ is defined to satisfy both $\varepsilon_{\rm{on}}^{(i+1)} - \varepsilon_{\rm{on}}^{(i)} = U$ and $\varepsilon_{\rm{on}}^{(1)} = -\varepsilon_{\rm{on}}^{(N)}$. We provide a detailed theoretical formulation for the $N=3$ case in Appendix~\ref{sec: Electronic structure of biased ATMG}, and further investigate the layer potentials of realistic dual-gating systems in Appendix~\ref{sec: Self-consistent Hartree calculation under dual-gating}. The low-energy effective Hamiltonian of each Dirac cone can be obtained from perturbation theory as \cite{Shin2023}
\begin{eqnarray} \label{eq: perturbative Hamiltonian}
 H_{\rm{eff}} = \Delta(\tilde{\alpha}) + \tilde{v} (\bm{k}\cdot \bm{\sigma}),   
\end{eqnarray}
\noindent where $\Delta(\tilde{\alpha})$ is the energy shift due to the potential difference $U$ and $\tilde{v}$ is the modified Fermi velocity of the Dirac cone. The gapped out-of-phase mode is generated from interband transitions between the shifted Dirac cones. As in AA-stacked graphene, an undamped out-of-phase plasmon appears when the chemical potential exceeds the energy level associated with the maximum shift of the Dirac cone \cite{Mohammadi2021}.
For $N=4$, the shifts at $\bar{K}$ and $\bar{K}'$ occur in opposite directions, which makes Landau damping occur more easily for the out-of-phase mode. Furthermore, for larger $N$ the increasing maximum shift hinders the formation of undamped out-of-phase modes \cite{Shin2023}. We therefore restrict our analysis to the $N=3$ case.

\begin{figure}[htb]
\includegraphics[width=1.0\linewidth]{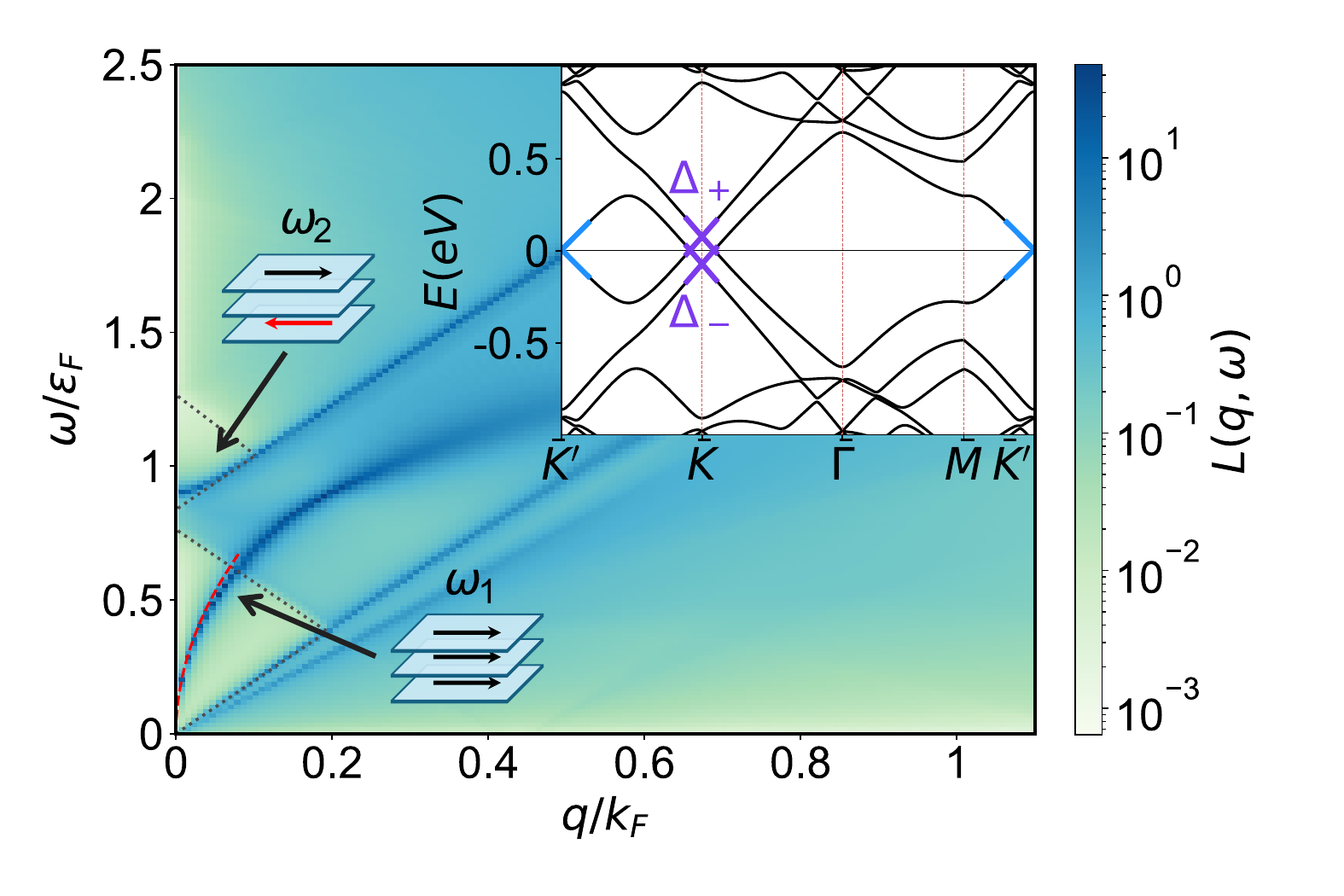}
\caption{Loss function $L(\bm{q}, \omega)$ of a biased AT3G system at a twist angle $\theta = 5^{\circ}$ under the interlayer potential difference $U = 0.08$ eV with $\varepsilon_{\text{F}} = 0.2$ eV ($n_{\rm{tot}} =1.4 \times 10^{13}$ cm$^{-2}$). The inset shows the band structure of a biased AT3G, where the colored lines highlight the perturbative Hamiltonian with energy splitting $\Delta_\pm$ near $\bar{K}$ and $\bar{K}'$.}\label{fig: fig4}
\end{figure}

At $\bar{K}$, the effective Hamiltonian of AT3G in the presence of an interlayer potential $U$ can be described as two Dirac cones with velocity $\tilde{v} = \frac{v_1 + v_2}{2}$, split in energy by $\Delta_\pm(\tilde{\alpha}) = \pm C(\tilde{\alpha})U$ with $C(\tilde{\alpha}) = \frac{1}{\sqrt{1+12\tilde{\alpha}^2}}$, whereas the effective Hamiltonian at $\bar{K}'$ remains unchanged by $U$ \cite{Shin2023}. Figure~\ref{fig: fig4} shows that the in-phase mode is generated from the two shifted Dirac cones at $\bar{K}$ and one Dirac cone at $\bar{K}'$, and it still exhibits a $\sqrt{q}$ dispersion. For the gapped out-of-phase mode, which arises from interband transitions between the shifted Dirac cones, the gap $\omega_{\rm{gap}, 2}$ can be roughly approximated as
\begin{eqnarray}
    \omega_{\rm{gap}, 2} \approx  2C(\tilde{\alpha}) U.
\end{eqnarray}
\noindent Note that the perturbation $U$ mixes the eigenvectors of the one-dimensional chain model, so the selection rules for the Coulomb eigenvectors may no longer be applicable. Indeed, as $U$ increases, the out-of-phase plasmon gap approaches the interband transition energy because $U$ suppresses the Coulomb-interaction-induced shift of the plasmon gap (see Appendix~\ref{sec: Electronic structure of biased ATMG}). Furthermore, at larger $U$, the mode eventually becomes Landau damped, no longer satisfying $\omega_{\rm{gap},2}<2[\varepsilon_{\rm F}-\Delta_+(\tilde{\alpha})]$ (which simplifies to $\omega_{\rm{gap},2} < \varepsilon_{\rm{F}}$) as in AA-stacked graphene. Nevertheless, the out-of-phase mode still arises from the interband transitions between the two Dirac cones, indicating that the gap can be tuned by an applied gate electric field. 

\section{\label{sec: Discussion}Discussion}

\begin{figure}[htb]
\includegraphics[width=1.0\linewidth]{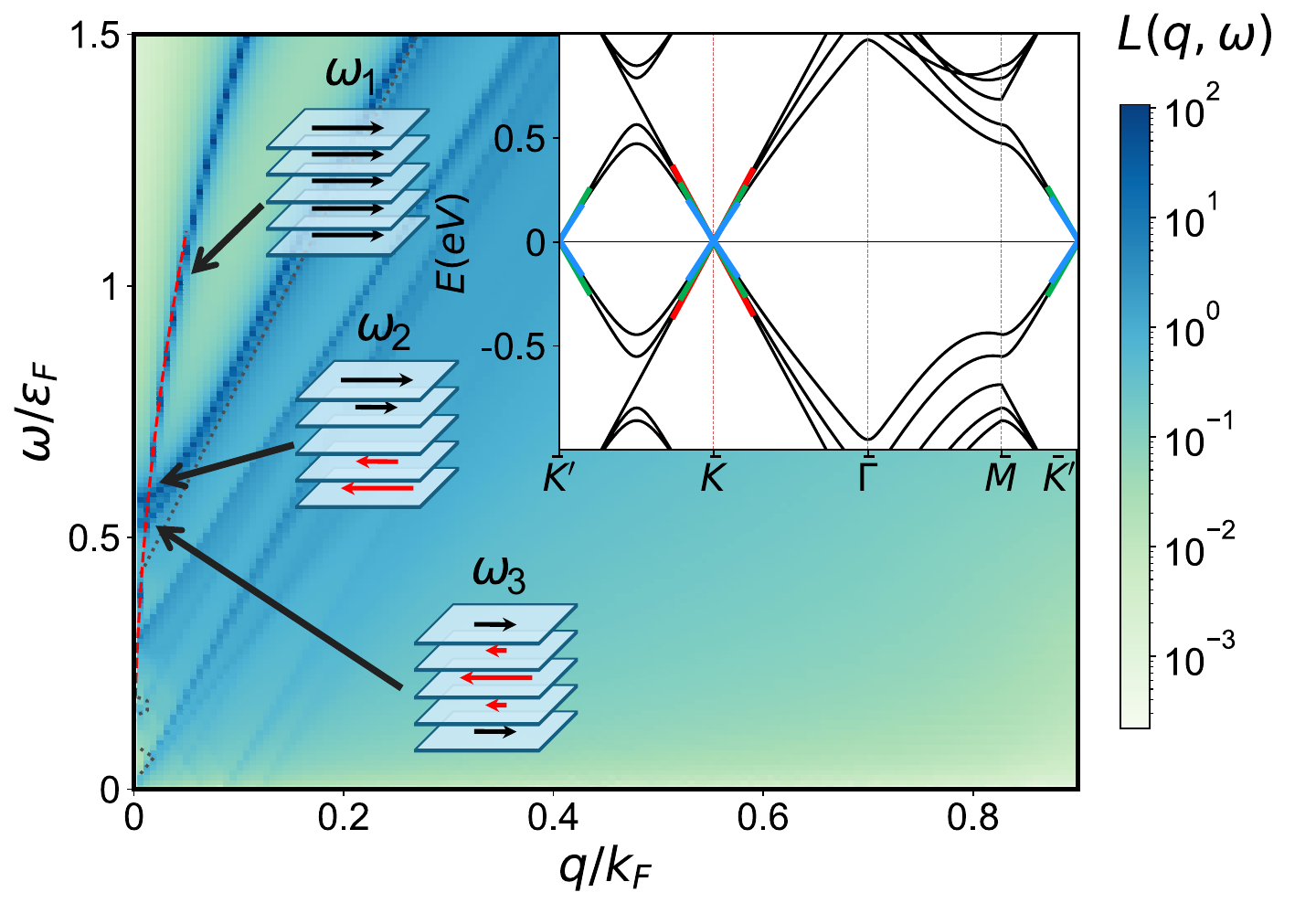}
\caption{Loss function $L(\bm{q}, \omega)$ of an unbiased AT5G system at a twist angle $\theta = 7^{\circ}$ with $\varepsilon_{\text{F}} = 0.3$ eV ($n_{\rm{tot}} = 4.1 \times 10^{13}$ cm$^{-2}$).
} \label{fig: fig5}
\end{figure}

The analytically derived out-of-phase plasmon gaps [Eqs. (\ref{eq: N3omega_gap_2}) and (\ref{eq: N4omega_gap_2})] for the effective Hamiltonian were obtained using the first-shell model, which employs nearest-neighbor truncation. When the twist angle decreases, the first-shell model is insufficient to accurately capture the band structure of the enlarged ATMG moir\'{e} superlattice, leading to discrepancies with the numerical results. In addition, as the carrier density increases, the low-energy effective Hamiltonian becomes less accurate at the Fermi level, so the approximately obtained plasmon gaps deviate from the numerical results. Nevertheless, our analytical analysis captures the out-of-phase plasmon mode in ATMG quite well in the large-twist-angle regime, consistent with full numerical calculations.

The present results focus on AT3G and AT4G. For the $N=5$ case, three Dirac cones labeled by $v_1, v_2,$ and $v_3=v_0$ exist near $\bar{K}$ and two Dirac cones labeled by $v_1$ and $v_2$ exist near $\bar{K}'$ as shown in Fig.~\ref{fig: fig5}. In contrast to the $N=3$ and $N=4$ cases, additional interband transitions between Dirac cones at $\bar{K}$ emerge, giving rise to the out-of-phase mode $\omega_3$ along with the out-of-phase mode $\omega_2$. As in the analysis above, intraband transitions generate the in-phase mode $\omega_1$, interband transitions between $v_1$ and $v_2$ and between $v_2$ and $v_3=v_0$ generate the out-of-phase mode $\omega_2$, and interband transitions between $v_1$ and $v_3=v_0$ generate the out-of-phase mode $\omega_3$. Note that the plasmon gaps exhibit the behavior $\omega_{{\rm gap},\alpha} \approx v_{\alpha 1} k_{{\rm F},1}$ in the weak Coulomb-interaction limit, which suggests that a larger interband transition scale $v_{\alpha 1} k_{{\rm F},1}$ leads to a larger plasmon gap. However, the numerical results show that $\omega_{\rm{gap},2}$ is larger than $\omega_{\rm{gap},3}$. This difference arises because $\omega_2$ is generated by multiple interband transitions at $\bar{K}$ and $\bar{K}'$, whereas $\omega_3$ is generated by a single interband transition at $\bar{K}$. As a result, the additional Coulomb-interaction contribution [see the additional term in Eq.~(\ref{eq: N3omega_gap_2}) for the $N=3$ case] becomes dominant for $\omega_2$, yielding a higher plasmon gap than that of $\omega_3$.

\begin{figure}[htb]
\includegraphics[width=1.0\linewidth]{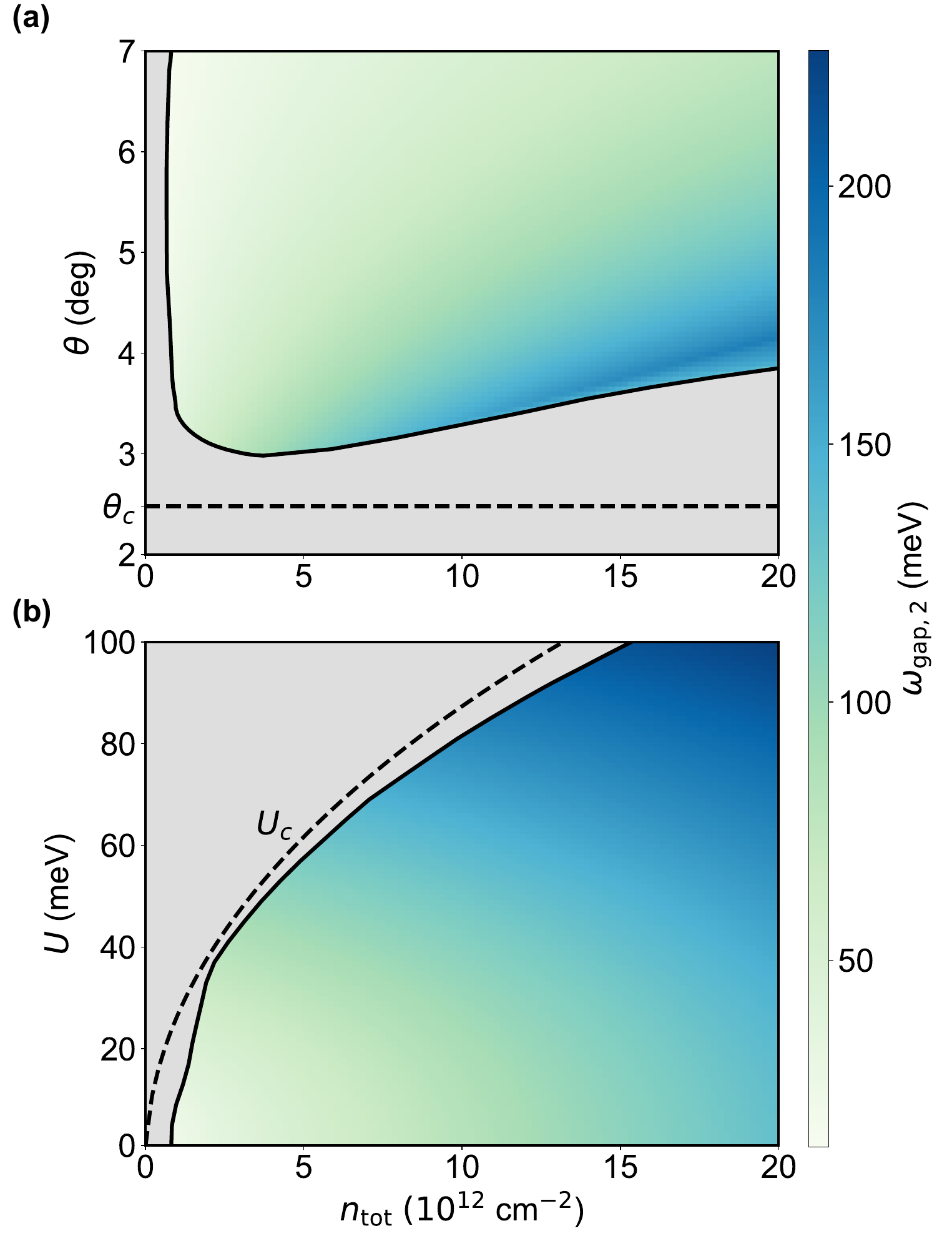}
\caption{The out-of-phase plasmon gap $\omega_{\rm{gap}, 2}$ of an AT3G system is plotted as a function of (a) $\theta$ and $n_{\rm{tot}}$ with $U=0$ eV and (b) $U$ and $n_{\rm{tot}}$ for an angle $\theta = 5^\circ$. The black solid lines indicate the numerically determined boundaries where the plasmon is undamped. The black dashed lines represent the analytically obtained critical lines, using $\theta_c$ from $v_\alpha = (1 + \sqrt{2})v_1$ with the largest Dirac velocity $v_\alpha$ in (a) and $U_c$ from $U = \frac{\varepsilon_{\text{F}}}{2C(\tilde{\alpha})}$ in (b), respectively.
}
\label{fig: fig6}
\end{figure}

In our calculations, we focus on representative twist angles in the range $5^\circ-7^\circ$, which lie above the typical magic-angle regime of $1^\circ-2^\circ$. Since the gapped out-of-phase mode arises from interband transitions between moir\'e Dirac cones of different velocities, its gap is given by $\omega_{\rm{gap},\alpha} \approx v_{\alpha1}k_{\rm{F},1} = (v_\alpha/v_1 - 1)\varepsilon_{\rm{F}}$, and grows with the carrier density $n_{\rm{tot}}$. As the twist angle decreases, all Dirac velocities are renormalized downward, but $v_1$ is suppressed the most rapidly. Therefore $\omega_{\rm{gap},\alpha}$ increases monotonically as $\theta$ decreases, until $v_\alpha < (1+\sqrt{2})v_1$ is no longer satisfied and the out-of-phase mode becomes damped. Figure~\ref{fig: fig6}(a) shows $\omega_{\rm gap,2}$ of an unbiased AT3G as a function of $\theta$ and $n_{\rm tot}$. As expected, an undamped gapped out-of-phase mode appears only for angles above $\theta_c = 2.75^\circ$, but the numerical results exhibit some deviations from the analytic prediction obtained from the low-energy effective Dirac Hamiltonian. At low angles, the density range where the low-energy effective model remains valid shrinks, and the damping region for $\omega_2$ expands with increasing density. At low densities, $\omega_2$ is damped due to the use of a finite broadening parameter $\eta$ rather than an infinitesimal one. For these reasons, we expect the out-of-phase modes to be completely suppressed by the particle-hole continuum of the flat bands below the critical angle, although more sophisticated calculations accounting for the Hartree potential and local-field corrections are required in the magic-angle regime \cite{Cavicchi2024}.

In the presence of a perpendicular electric field, the analytic form of the effective perturbative Hamiltonian in Eq.~(\ref{eq: perturbative Hamiltonian}) is valid within a radius of about $k_c \sim U/v_0$ in momentum space. Within
this range, the out-of-phase mode in AT3G arises from interband transitions between the
two Dirac cones split by $\Delta_\pm = \pm C(\tilde{\alpha})U$, so that
$\omega_{\rm{gap},2} \approx 2C(\tilde{\alpha})U$. The gap therefore increases
monotonically as $U$ increases, until $\omega_{\rm{gap},2} < \varepsilon_{\rm{F}}$
is no longer satisfied and the out-of-phase mode becomes damped. Figure \ref{fig: fig6}(b) illustrates $\omega_{\rm{gap}, 2}$ in AT3G as a function of $U$ and $n_{\rm{tot}}$ at $\theta = 5^\circ$. The result demonstrates that the out-of-phase mode is undamped if the potential difference $U$ satisfies $2C(\tilde{\alpha}) U< \varepsilon_{\text{F}}$ in the weak Coulomb-interaction limit. We analytically obtain the critical density using $U_c$ which corresponds to the black dashed line in Fig.~\ref{fig: fig6}(b). As in Fig.~\ref{fig: fig6}(a), the numerical results exhibit deviations arising from a finite broadening parameter $\eta$ at low densities and from the validity of the first-order correction in $U$ at high densities.

We point out that ATMG provides a robust and versatile multilayer-graphene platform for engineering an undamped gapped out-of-phase plasmon mode at experimentally reasonable carrier densities. In contrast to AA-stacked (AB-stacked) bilayer graphene, where accessing an undamped gapped out-of-phase mode typically requires the Fermi energy $\varepsilon_{\rm{F}} \gtrsim t_{\perp}$ ($\varepsilon_{\rm{F}} \gtrsim t_\perp/2$) with the interlayer tunneling $t_\perp$ \cite{Borghi2009, Mohammadi2021}, corresponding to a high carrier density $n_{\rm{tot}}\sim 10^{13}$ cm$^{-2}$, the out-of-phase mode in ATMG, in principle, remains undamped regardless of $n_{\rm{tot}}$ in the large-twist-angle regime. Moreover, the plasmon spectrum in ATMG is highly tunable through the twist angle, carrier doping, and an external electric field. It may be possible to measure this gapped out-of-phase plasmon mode via electron energy loss spectroscopy (EELS), scattering-type scanning near-field optical microscopy (s-SNOM), or on-chip photocurrent measurements \cite{Hesp2021, Huang2022, Politano2011, Hu2017, Menabde2021, Ruiz2023}.

We also perform self-consistent Hartree calculations to confirm that the gapped out-of-phase modes arise under realistic dual-gating systems. In such configurations, the layer-dependent Hartree potentials become symmetric without an external electric field, while they exhibit asymmetry under an applied field. We show that these effects act as an interlayer potential difference and can be treated perturbatively, modifying only the gapped out-of-phase plasmon modes quantitatively without changing the underlying physics for $N=3$ and $N=4$ in the main text. We leave the details in Appendix~\ref{sec: Self-consistent Hartree calculation under dual-gating}.

In summary, we have investigated plasmons in ATMG both analytically and numerically. Using a continuum Hamiltonian, we obtained numerical results for the plasmon modes, whereas using the low-energy effective Hamiltonian and the Coulomb eigenvector basis, we obtained the corresponding analytical results. The in-phase mode exhibits a square-root dispersion of the form $\sqrt{q/\varepsilon_{\rm{out}}}$ with the dielectric constant $\varepsilon_{\rm{out}}$, while the out-of-phase plasmon modes develop gaps governed by specific interband transitions between Dirac cones with the dielectric constant ${\varepsilon_{\rm{in}}}$. We also demonstrated that, in the weak Coulomb-interaction limit, these out-of-phase modes are undamped when the twist angle is above the critical value, regardless of the density as long as the low-energy effective Dirac Hamiltonian remains valid. We also perform self-consistent Hartree calculations to confirm that the gapped out-of-phase modes arise under realistic dual-gating systems. Furthermore, we found that the gap of the out-of-phase mode can be tuned by applying a perpendicular electric field, which shifts the Dirac cones and gives rise to plasmon features reminiscent of those observed in AA-stacked multilayer graphene. The analytic and numerical results for ATMG obtained here will help clarify the plasmonics of moir\'e systems.

\acknowledgments
The work at SNU was supported by the National Research Foundation of Korea (NRF) grants funded by the Korea government (MSIT) (Grant No. RS-2023-NR076715), the Creative-Pioneering Researchers Program through Seoul National University (SNU), and the Center for Theoretical Physics.

\appendix

\section{Multilayer Coulomb interaction} \label{sec: Multilayer Coulomb interaction}

\begin{figure}[htb]
\includegraphics[width=0.8\linewidth]{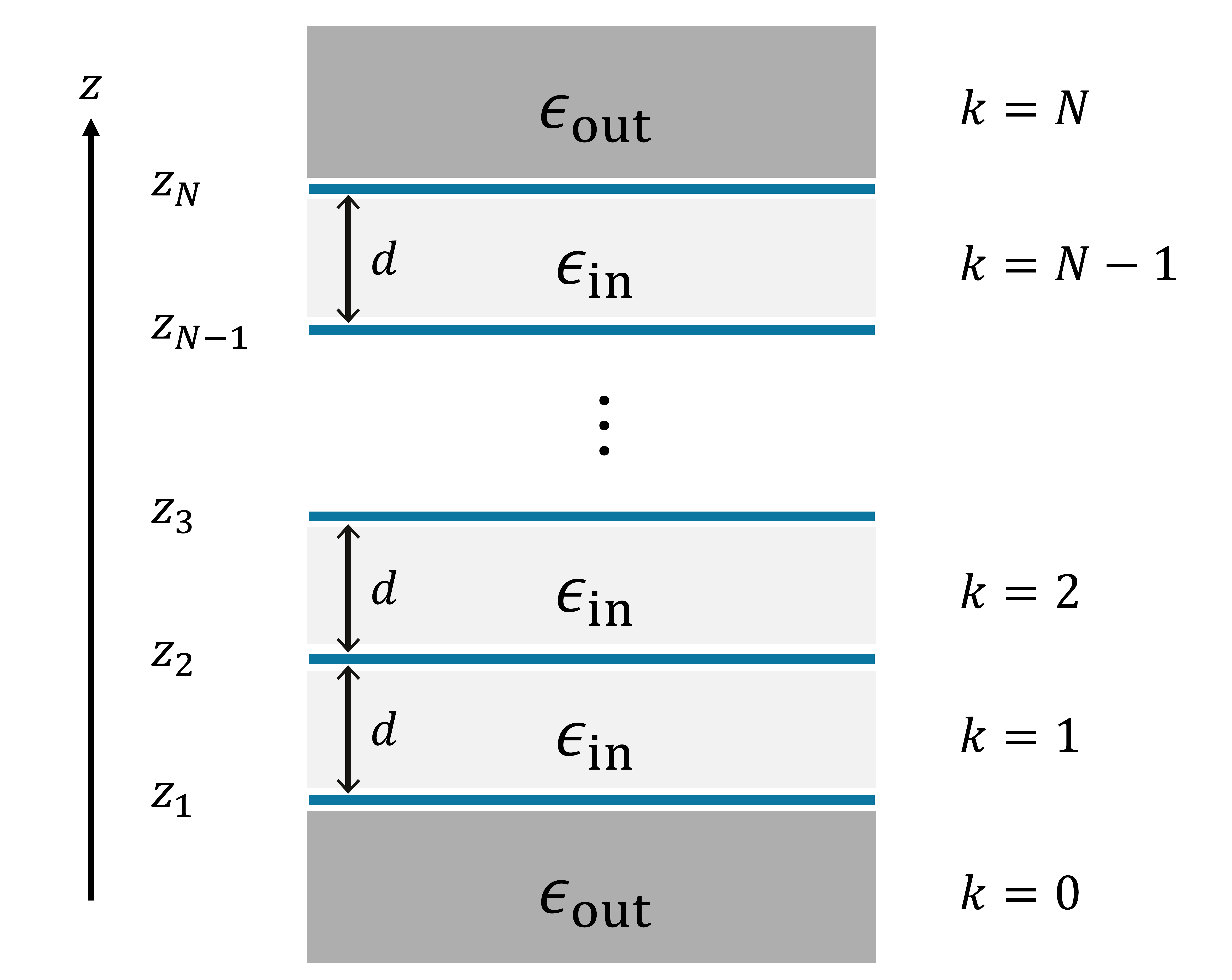}
\caption{Schematic illustration of the dielectric environment, consisting of $N$ equally spaced sheets separated by a distance $d$. The system is characterized by dielectric constants $\epsilon_{\rm in}$ (inside) and $\epsilon_{\rm out}$ (outside). Here, $z$ denotes the out-of-plane direction, where the sheet locations are given by $z_k = (k-1)d$ with $k = 1, 2, \dots, N$, and the regions are indexed by $k=0, 1, \dots, N$.
}\label{fig: fig7}
\end{figure}

We consider a system consisting of $N$ equally spaced sheets separated by a distance $d$ as shown in Fig.~\ref{fig: fig7}. The sheets are located at $z_k = (k-1)d$ with $k = 1, 2, \dots, N$.  The bare Coulomb matrix $V_{ij}(q)$ in the $N$-layer system can be determined by solving the Poisson equation \cite{Profumo2010}
\begin{eqnarray}
    \nabla\cdot[\epsilon(z) \nabla\phi(\bm{\rho}, z)] = -4\pi \sigma_j,
\end{eqnarray}
where $\phi(\bm{\rho}, z)$ denotes the electrostatic potential induced by a point charge $\sigma_j = (-e) \delta(\bm{\rho})\delta(z-z_j)$ in cylindrical coordinates. Here, the dielectric function is taken to be $\epsilon(z) = \epsilon_{\rm out}$ for the outer regions ($z < z_1$ or $z>z_N$), and $\epsilon(z) =\epsilon_{\rm in}$ for the inner regions ($z_{k}<z<z_{k+1}$). For convenience, we introduce an additional labeling such that the regions are indexed by $k = 0, 1, \dots, N$: regions $k = 0$ and $k=N$ correspond to the outer regions while $k=1, 2, \dots, N-1$ denote the inner regions. The electric potential $\phi(\bm{\rho}, z)$ can be defined piecewise in each of the $N+1$ regions. In the $k$th region, the electric potential takes the form
\begin{equation}
 \phi ( \bm{\rho}, z) = \int \frac{d^2 q}{(2\pi)^2}e^{i\bm{q}\cdot\bm{\rho}} (A_k e^{qz} + B_k e^{-qz}),
\end{equation}
\noindent which involves a total of $2(N+1)$ constants $A_k$ and $B_k$ with the in-plane momentum $\bm{q}$ and its magnitude $q = |\bm{q}|$. Note that the boundary conditions at infinity, $\phi(\bm{\rho}, z \to \pm \infty)=0$, impose $B_0=0$ and $A_N=0$. The remaining $2N$ coefficients are determined by enforcing the continuity of the potential $\phi(\bm{\rho}, z)$ and the discontinuity in the $z$ component of the electric displacement $\bm{D}(\bm{\rho}, z) = \epsilon(z) \bm{E}(\bm{\rho}, z)$ with the electric field $\bm{E} = -\nabla\phi$.

After solving a routine electrostatic calculation, the Coulomb matrix $V_{ij}(q) = (-e) \tilde{\phi}(\bm{q}, z = z_i)$, where $\tilde{\phi}(\bm{q}, z)$ denotes the 2D Fourier transform of the electric potential induced by $\sigma_j$, can be obtained as
\begin{eqnarray} \label{eq: full Coulomb interaction}
    V_{ij}(q) &=& \frac{2\pi e^2}{ \epsilon_{\rm{in}}q} \frac{1}{f(q)} [\epsilon_+^2 e^{x_{ij}qd} + \epsilon_-^2 e^{-x_{ij}qd } \nonumber \\
    &+& 2\epsilon_+ \epsilon_- \cosh{(y_{ij}qd)}],
\end{eqnarray}
\noindent where $f(q) = \epsilon_+^2 e^{q(N-1)d} - \epsilon_-^2 e^{-q(N-1)d}$ with $\epsilon_{\pm} = \epsilon_{\rm{in}} \pm \epsilon_{\rm{out}}$. Here, we define $x_{ij} = N-|i-j|-1$ and $y_{ij} = N - i - j +1$ for simplicity. If $\varepsilon_{\rm{in}} = \varepsilon_{\rm{out}}$, the interlayer Coulomb interaction reduces to a trivial form, $V_{ij} = \frac{2\pi e^2}{\epsilon_{\rm{in}}q}e^{-q|i-j|d}$. In the long-wavelength limit ($q\rightarrow 0$), the Coulomb matrix in Eq.~(\ref{eq: full Coulomb interaction}) becomes
\begin{eqnarray}\label{eq: Coulomb interaction}
    V_{ij}(q) &=& \frac{8\pi e^2}{q}\frac{\epsilon_{\rm{in}}}{f(q)}\bigg[1 + \frac{\epsilon_{\rm{out}}}{\epsilon_{\rm{in}}}x_{ij}qd + O(q^2)\bigg]   \\
    &=& \frac{8\pi e^2}{q}\frac{\epsilon_{\rm{in}}}{f(q)}\bigg[\exp{\bigg({\frac{\epsilon_{\rm{out}}}{\epsilon_{\rm{in}}}}x_{ij}qd\bigg)} + O(q^2)\bigg], \nonumber
\end{eqnarray}
\noindent where we use $\exp(x) \approx 1 + x$. Note that the Coulomb matrix in the long-wavelength limit is proportional to the KMS matrix $A_{ij}(\rho) = \rho^{|i-j|}$ with $\rho = \exp{(-\frac{\epsilon_{\rm{out}}}{\epsilon_{\rm{in}}}qd)}$ \cite{Kac1953, Trench2001, Bogoya2016, Fikioris2018, Narayan2021, Kim2025}. By using the eigenvalues and eigenvectors of the KMS matrix, we can obtain Eq.~(\ref{eq:rho->1 limit}).

When the top gate and bottom gate are introduced, the boundary conditions are modified by $\phi(\bm{\rho}, z= -d_g / 2) = \phi(\bm{\rho}, z = (N-1)d + \frac{d_g}{2})$. Solving the electrostatic calculation under these conditions, we obtain
\begin{eqnarray}
    V_{ij}(q) &=& \frac{2\pi e^2}{\epsilon_{\rm{in}} q} \frac{1}{g(q)} [K_+^2 e^{x_{ij} qd} + K_-^{2} e^{-x_{ij}qd} \nonumber \\
    &+& 2K_+ K_- \cosh(y_{ij}qd)],
\end{eqnarray}
where $g(q) = K_+^2 e^{q(N-1)d} - K_-^2 e^{-q(N-1)d}$ with $K_\pm = \epsilon_{\rm{in}}\pm \epsilon_{\rm{out}}\coth(q\frac{d_g}{2})$. Note that as $d_g \rightarrow \infty$, $K_\pm$ becomes $\epsilon_\pm$, which is consistent with Eq.~\eqref{eq: full Coulomb interaction}. In the long-wavelength limit, we have
\begin{eqnarray}
    V_{ij}(q) &=& \frac{8\pi e^2}{q}\frac{\epsilon_{\rm{in}}}{g(q)}\bigg[1 + \frac{\epsilon_{\rm{out}}}{\epsilon_{\rm{in}}}\frac{2d}{d_g}x_{ij} + \mathcal{O}(q)\bigg] \\
    &=& \frac{8\pi e^2}{q}\frac{\epsilon_{\rm{in}}}{g(q)} \bigg[\exp\bigg(\frac{\epsilon_{\rm{out}}}{\epsilon_{\rm{in}}}\frac{2d}{d_g} x_{ij}\bigg) + \mathcal{O}(q) \bigg]. \nonumber
\end{eqnarray}
This demonstrates that in the long-wavelength limit, the Coulomb interaction incorporating the screening effects is proportional to the KMS matrix $A_{ij} = \rho^{|i-j|}$, where $\rho = \exp(-\frac{\epsilon_{\rm{out}}}{\epsilon_{\rm{in}}}\frac{2d}{d_g})$. Thus, the essential structure of the KMS matrix is preserved even with metallic gates. Since the presence of these gates only slightly alters our results quantitatively under typical experimental conditions ($\frac{d_g}{2} \sim 10$ nm and $-\ln\rho\sim \frac{3}{100}\frac{\epsilon_{\rm{out}}}{\epsilon_{\rm{in}}}$), we focus on the $d_g \rightarrow \infty$ case throughout the paper.

\section{Wave functions of ATMG}\label{sec: Wave functions of ATMG}

Following \cite{Khalaf2019, Leconte2022, Shin2023}, we construct the wave functions of ATMG at $\bar{K}$ or $\bar{K}'$ using the first-shell model, where the momentum-space lattice consists of the nearest-neighbor shell of moir\'e reciprocal lattice vectors under the rigid model ($w = w'$). Near $\bar{K}$, the term $\psi_{r, s}$ in Eq.~\eqref{eq: egienvector of ATMG} denotes the eigenfunction components of ATMG, which are constructed from the normalized eigenstates $a_s$ of $\bm{k}\cdot\bm{\sigma}_{\theta/2}$ as
\begin{eqnarray}\label{eq: componenets of wavefunction}
    \psi_{r, s} =\begin{cases}
        a_s, &  r \text{ odd}, \\
        t_r b_s , &  r \text{ even},
    \end{cases}
\end{eqnarray}
\noindent where $b_s = (b_{\bm{q}_0, s}, b_{\bm{q}_+, s}, b_{\bm{q}_-, s})^T$ is determined by $b_{\bm{q}_j, s} = -[ v_0 (\bm{k} + \bm{q}_j)\cdot \bm{\sigma}_{\theta/2}]^{-1} T_j^{\dagger} a_{s}$ with $t_r = 2\cos{\phi_r}$. Near $\bar{K}'$, the roles are exchanged, so that $\psi_{r,s} = a_s$ for even $r$, while $\psi_{r,s} = t_r b_s$ for odd $r$.

For the $N=3$ case, using Eqs.~(\ref{eq: egienvector of ATMG}) and (\ref{eq: componenets of wavefunction}), we obtain the normalized wave functions $|\Psi_{r, s}\rangle$ near $\bar{K}$ within the first-shell model as
\begin{subequations}\label{eq: wave functions of AT3G at K}
    \begin{flalign}
    |\Psi_{1, s}\rangle &= \frac{1}{\sqrt{2 + 24\tilde{\alpha}^2}}\begin{pmatrix}
        a_s \\ 2b_{s} \\ a_s
    \end{pmatrix}, \\
    |\Psi_{2, s}\rangle &= \frac{1}{\sqrt{2}} \begin{pmatrix}
        a_s \\ 0 \\ -a_s
    \end{pmatrix},
    \end{flalign}
\end{subequations}
\noindent and near $\bar{K}'$ as
\begin{eqnarray}\label{eq: wave functions of AT3G at K'}
    |\Psi_{1, s}\rangle = \frac{1}{\sqrt{1+12\tilde{\alpha}^2}} \begin{pmatrix}
        b_s \\ a_{s} \\ b_s
    \end{pmatrix}.
\end{eqnarray}
\noindent Note that the wave functions associated with the $v_1$ Dirac cones at both $\bar{K}$ and $\bar{K}'$ exhibit symmetric behavior, whereas the wave function of $v_2 = v_0$ at $\bar{K}$ exhibits antisymmetric behavior. 

Similarly, for the $N=4$ case, one can derive the wave functions of Dirac cones near $\bar{K}$ as
\begin{eqnarray}\label{eq: wave functions of AT4G at K}
    |\Psi_{r, s}\rangle = \frac{2}{\sqrt{5(1+6 t_r^2 \tilde{\alpha}^2)}}\begin{pmatrix}
    \sin{\phi_r} \cdot a_s \\
    \sin{2\phi_r} \cdot t_r b_s \\
    \sin{3\phi_r} \cdot a_s \\
    \sin{4\phi_r} \cdot t_r b_s 
    \end{pmatrix},
\end{eqnarray}
\noindent and near $\bar{K}'$ as
\begin{eqnarray}\label{eq: wave functions of AT4G at K'}
    |\Psi_{r, s}\rangle = \frac{2}{\sqrt{5(1+6 t_r^2 \tilde{\alpha}^2)}}\begin{pmatrix}
    \sin{\phi_r} \cdot t_r b_s \\
    \sin{2\phi_r} \cdot a_s \\
    \sin{3\phi_r} \cdot t_r b_s \\
    \sin{4\phi_r} \cdot a_s 
    \end{pmatrix}.
\end{eqnarray}
\noindent As in the $N=3$ case, the $v_1$ wave functions are symmetric at both $\bar{K}$ and $\bar{K}'$, in contrast to the antisymmetric nature of $v_2$ wave functions at both $\bar{K}$ and $\bar{K}'$. 

\section{Calculation of the noninteracting density-density response function}\label{sec: Calculation of the noninteracting density-density response function}

We calculate the noninteracting density-density response function in the Coulomb eigenvector basis within the first-shell model in the high-angle regime. In this limit, the velocities of the Dirac cones $v_r$ approach $v_0$ since $\tilde{\alpha} = \frac{w}{2 v_0 k_{\rm{D}} \sin{(\theta / 2)}} \ll 1$, which holds for $\theta \geq 5^\circ$. This implies that the system is effectively described by decoupled Dirac cones, corresponding to the weak-tunneling limit. Furthermore, given the block-diagonal nature of $\chi_{\alpha\beta}^{(0)}$ [$\chi^{(0)} = \chi_{\text{odd}}^{(0)} \oplus \chi_{\text{even}}^{(0)}$, where the block $\chi_{\text{odd}(\text{even})}^{(0)}$ consists of elements with odd (even) indices] and the correspondence of the wave functions to the one-dimensional chain model solutions, the plasmon modes are approximately determined by $1-V_\alpha \chi_{\alpha\alpha}^{(0)} = 0$ \cite{Kim2025}. In the following, we focus on the cases $N=3$ and $N=4$. Note that we only need to consider $\alpha = 1, 2$ because there are at most two Dirac cones at both the $\bar{K}$ and $\bar{K}'$ valleys. 

In the long-wavelength limit, the in-phase mode $\omega_1$ is determined by $1 - V_1(q)\chi^{(0)}_{11}(\bm{q}, \omega) = 0$. Clearly, the noninteracting response function $\chi_{11}^{(0)}$ consists of the intraband contributions of Dirac cones and is expressed as
\begin{eqnarray}\label{eq: intraband contribution of response function}
    \chi_{11}^{(0)}(\bm{q}\rightarrow 0, \omega) = \sum_{\nu, r} \frac{n_{\nu, r}}{m_r} \frac{q^2}{\omega^2}
\end{eqnarray}
where $m_r = k_{\text{F}, r} / v_r$ and $n_{\nu, r} = g k_{\text{F}, r}^2 / 4\pi$. Using Eqs.~(\ref{eq:rho->1 limit}a) and (\ref{eq: intraband contribution of response function}), we obtain the in-phase mode dispersion in Eq.~(\ref{eq: in-phase dispersion and out-of-phase mode}a).

For the out-of-phase mode $\omega_2$, we obtain the noninteracting response function $\chi_{22}^{(0)}$. Since the plasmon dispersion is determined by $1 - V_2 \chi^{(0)}_{22} = 0$, the plasmon mode requires $\chi^{(0)}_{22}(q, \omega)$ to be real and positive. Consequently, we restrict our analysis to the region outside the particle-hole continuum and exclude regimes where $\rm{Re}[\chi_{22}^{(0)}] < 0$. The response function in the zero-momentum limit is governed solely by interband transitions between $|\Psi_{1, s}\rangle$ and $|\Psi_{2, s}\rangle$, given by
\begin{eqnarray}\label{eq: out-of-phase density response function}
\chi^{(0)}_{22}(q\rightarrow 0, \omega) &=& g \sum_{\nu ,ss', r\neq r'}\int\frac{d^2k}{(2\pi)^2}\frac{f_{\bm{k}, r, s} - f_{\bm{k}, r', s'}}{\omega + (sv_{r}-s'v_{ r'})k} \nonumber \\
&\times& |\langle\Psi_{r, s}|U_{2}|\Psi_{r', s'}\rangle|^2 \nonumber \\
&=& g F_N\int_{k_{\rm{F}, 2}}^{k_{\rm{F}, 1}}\frac{kd k}{2\pi}\bigg[\frac{1}{\omega - v_{21}k} - \frac{1}{\omega + v_{21}k}\bigg] \nonumber \\
&=& g F_N\frac{\omega}{2\pi v_{21}^2}\bigg[2\frac{v_{21}}{\omega}(k_{\rm{F}, 2}-k_{\rm{F}, 1}) \nonumber \\
&+& \ln{\frac{(\omega + v_{21}k_{\rm{F, 1}})(\omega - v_{21}k_{\rm{F,2}})}{(\omega - v_{21}k_{\rm{F, 1}})(\omega + v_{21}k_{\rm{F,2}})}}\bigg].
\end{eqnarray}
\noindent which is valid for $\omega > v_{21} k_{\text{F}, 1}$. Here, we used $|\Psi_{1,s}\rangle$ and $|\Psi_{2,s}\rangle$ obtained in Appendix~\ref{sec: Wave functions of ATMG}, and $F_N=|\langle \Psi_{1, s}|U_2|\Psi_{2, s}\rangle|^2$ represents the overlap factor with the antisymmetric oscillation matrix $U_2$. For $N=3$ and $N=4$, these factors are explicitly given as
\begin{subequations}\label{eq:rho->1 limit}
    \begin{flalign}
    F_3 &= \frac{1}{2+24\tilde{\alpha}^2}, \\
    F_4 &= \frac{(2+\sqrt{2})(1+6t_1t_2\tilde{\alpha}^2)^2}{20(1+6 t_1^2 \tilde{\alpha}^2)(1+6t_2^2\tilde{\alpha}^2)}.
    \end{flalign}
\end{subequations}
\noindent To demonstrate that the intraband contribution to $\chi_{22}^{(0)}$ vanishes, we evaluate the element $\langle \Psi_{r, s}|U_2|\Psi_{r, s}\rangle$. For the $N=3$ case, it is straightforward to see that this element is zero. Note that, unlike in the $N=3$ case, this term does not vanish at each valley for $N=4$; instead, the contributions from $\bar{K}$ and $\bar{K}'$ mutually cancel. In the zero-momentum limit, the equation $1 - V_2(q)\chi^{(0)}_{22}(\bm{q}, \omega) = 0$ reduces to the following expression for the out-of-phase plasmon gap:
\begin{eqnarray}
    \omega = v_{21}k_{\text{F}, 1} + \delta\omega_2^{(N)}
\end{eqnarray}
\noindent where
\begin{eqnarray}
    & &\delta\omega_2^{(N)} = \frac{(\omega + v_{21}k_{\rm{F}, 1})(\omega - v_{21}k_{\rm{F}, 2})}{\omega + v_{21}k_{\rm{F}, 2}}  \\
    &\times&\exp{\bigg[\frac{v_{21}}{\omega}}  \bigg\{2(k_{\rm{F}, 2} - k_{\rm{F}, 1}) 
    - \frac{v_{21}\varepsilon_{\rm{in}}(1-\cos{(\frac{\pi}{N})})}{F_Ne^2 gd}\bigg\} \bigg]. \nonumber
\end{eqnarray}
\noindent Clearly, $\delta\omega_{2}^{(N)} \to 0$ when $\frac{e^2}{v_{21} \varepsilon_{\rm{in}}} \to 0$, meaning that the out-of-phase plasmon gap becomes $\omega_{\rm{gap}, 2} = v_{21}k_{\rm{F}, 1}$ in the weak Coulomb-interaction limit. Note that while Eq.~(\ref{eq: out-of-phase density response function}) presents several interband transitions between the $v_2$ and $v_1$ Dirac cones, the dominant contribution among the nonvanishing terms arises from the transition associated with $v_{2 1} k_{\rm{F}, 1}$.

The analysis presented above for $N=3$ and $N=4$ can be generalized to higher $N$ systems. In the presence of multiple Dirac cones, the dominant contribution in $\chi_{\alpha\alpha}^{(0)}$ arises from transitions between Dirac cones whose indices differ by $\alpha - 1$ \cite{Kim2025}. Since the high-angle regime corresponds to the weak tunneling limit, the several interband transitions for the out-of-phase mode $\omega_\alpha$ can be approximately replaced by the transitions between the $v_\alpha$ and $v_1$ Dirac cones, analogous to the isotropic tunneling case \cite{Kim2025}. This implies that the out-of-phase mode is described by Eq.~(\ref{eq: approximated form}) in the weak Coulomb-interaction regime.

\begin{figure}[htb]
\includegraphics[width=1.0\linewidth]{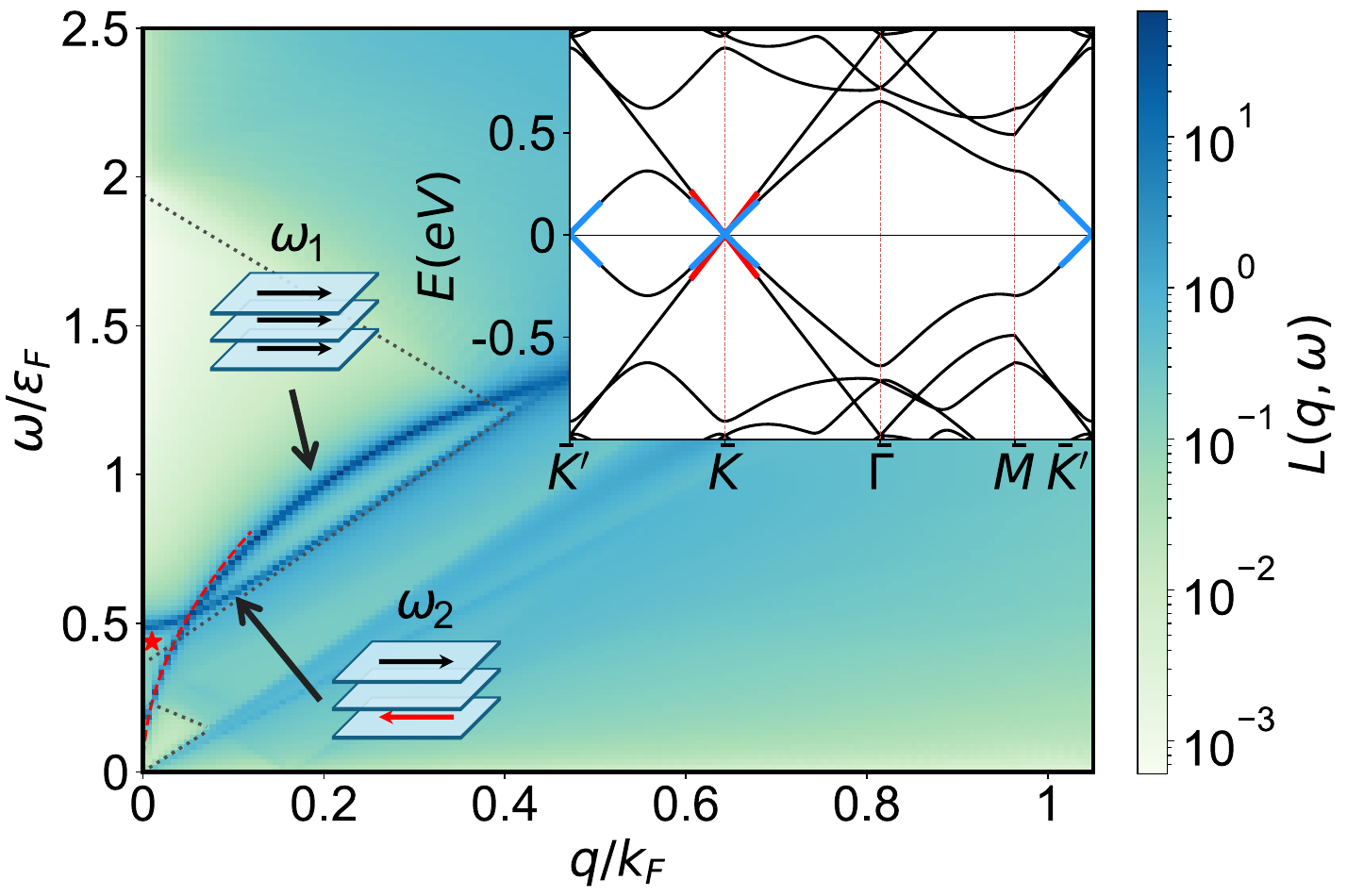}
\caption{Loss function $L(\bm{q}, \omega)$ of an unbiased AT3G system at a twist angle $\theta = 5^{\circ}$ with $\varepsilon_{\text{F}} = 0.2$ eV ($n_{\rm{tot}} = 1.3 \times 10^{13}$ cm$^{-2}$). Here, $\bm{q}$ is chosen along the $\bar{\Gamma}-\bar{K}$ direction.
}\label{fig: fig8}
\end{figure}

\section{Directional dependence of plasmon dispersion on momentum transfer $q$}\label{sec: Directional dependence of plasmon dispersion on momentum transfer}

We presented the plasmon dispersion along the $\bar{\Gamma}-\bar{M}$ direction in the main text, given that the low-energy Hamiltonian is effectively isotropic near the Dirac points. While the numerical band structure is anisotropic, this directional dependence becomes negligible for the long-wavelength behavior. Figure~\ref{fig: fig8} shows the loss function along the $\bar{\Gamma}-\bar{K}$ direction. This is consistent with that along the $\bar{\Gamma}-\bar{M}$ direction, confirming the isotropic character of the plasmon modes in the long-wavelength limit.

\section{Electronic structure of biased ATMG}\label{sec: Electronic structure of biased ATMG}

Following the perturbation theory of Shin \textit{et al.} \cite{Shin2023}, we consider the Hamiltonian for $N=3$ in the presence of the interlayer potential difference $U$. At $\bar{K}$, we calculate the perturbation matrix in the basis of the wave functions given in Eq.~(\ref{eq: wave functions of AT3G at K}) with $H_{\rm{on}} = \text{diag}[-U\mathbb{I}_2, \mathbf{0}_6, U\mathbb{I}_2]$. The $2 \times 2$ perturbation matrix elements are $[H_{\rm{on}}]_{11} = [H_{\rm{on}}]_{22} = 0$ and $[H_{\rm{on}}]_{12} = [H_{\rm{on}}]_{21} = - U /\sqrt{1 + 12\tilde{\alpha}^2}$, and the effective Hamiltonian for $N=3$ in Eq.~(\ref{eq: perturbative Hamiltonian}) is derived by diagonalizing the following matrix
\begin{eqnarray}\label{eq: perturbation theory AT3G}
    H_{\bar{K}} = \begin{bmatrix}
        v_1 (\bm{k}\cdot \bm{\sigma}) & -C(\tilde{\alpha})U  \\
        -C(\tilde{\alpha}) U  & v_2 (\bm{k}\cdot \bm{\sigma})
    \end{bmatrix},
\end{eqnarray}
\noindent where $C(\tilde{\alpha}) = 1 / \sqrt{1 + 12\tilde{\alpha}^2} $, assuming the regime $v_{21}k \ll U$ near $\bar{K}$. Note that the perturbed wave functions become linear combinations of $|\Psi_{1, s}\rangle$ and $|\Psi_{2, s}\rangle$. This mixing generates non-contributing components with vanishing elements ($\langle\Psi_{r, s}|U_2|\Psi_{r, s}\rangle = 0$) into the interband transitions between Dirac cones, leading to a reduction in the overlap factor $F_N$. Consequently, the Coulomb-interaction-induced shift of the plasmon gap is suppressed.

In contrast, at $\bar{K}'$, only a single Dirac cone with velocity $v_1$ exists, described by the wave function in Eq.~(\ref{eq: wave functions of AT3G at K'}). Consequently, the perturbation matrix vanishes, leaving the Dirac cone at $\bar{K}'$ unaltered to leading order in $U$. This implies that the out-of-phase mode $\omega_2$ is solely determined by the interband transitions at $\bar{K}$, with no contribution from the $\bar{K}'$ valley.

\section{Self-consistent Hartree calculations under dual-gating}\label{sec: Self-consistent Hartree calculation under dual-gating}

When ATMG is located between top and bottom metallic gates, charge carriers are redistributed, inducing an interlayer potential difference. We employ a self-consistent Hartree method to determine the electrostatic potential generated by these induced charges. Within the mean-field Hartree approximation, the Coulomb interaction can be reduced to a diagonal matrix $H_{\rm{H}} = \text{diag}[\varepsilon_{\rm{H}}^{(1)}, \varepsilon_{\rm{H}}^{(2)}, ..., \varepsilon_{\rm{H}}^{(N)}]$ representing the Hartree energy on each layer, where
\begin{eqnarray}
    \varepsilon_{\rm{H}}^{(i)} = \sum_{j}V_{ij}(0) n^{(j)}.
\end{eqnarray}
\noindent Here, $V_{ij}(0)$ is the Coulomb interaction defined in \eqref{eq: Coulomb interaction} and $n^{(i)}$ is the layer density, which is expressed as
\begin{eqnarray}
    n^{(i)} = \frac{g}{V}\sum_{\bm{k}, \lambda}f(\varepsilon_{\bm{k}, \lambda}) \langle\bm{k}, \lambda| P_i | \bm{k}, \lambda\rangle.
\end{eqnarray}
\noindent Then, the induced potential difference $\varepsilon_{\rm{H}}^{(ij)}=\varepsilon_{\rm{H}}^{(i)}-\varepsilon_{\rm{H}}^{(j)}$ between the $i$th layer and $j$th layer is given by \cite{Min2007, Jang2019, Kim2025b}
\begin{eqnarray}
    \varepsilon_{\rm{H}}^{(ij)}  &=& \sum_{k}[V_{ik}(0) - V_{jk}(0)]n^{(k)} \nonumber \\
    &=&\sum_k \frac{8\pi e^2}{q} \frac{\epsilon_{\rm{out}}}{\epsilon_+^2 - \epsilon_-^2}qd(x_{ik} - x_{jk}) n^{(k)} \nonumber \\
    &=& \sum_k \frac{2\pi e^2}{\epsilon_{\rm{in}} } (d_{ik} - d_{jk}) n^{(k)},
\end{eqnarray}
\noindent where $d_{ij} = |i-j|d$ is the perpendicular distance between the two layers. Therefore, the total onsite energy difference $\varepsilon_{\rm{on}}^{(ij)}=\varepsilon_{\rm{on}}^{(i)}-\varepsilon_{\rm{on}}^{(j)}$ between the $i$th and $j$th layers including the Hartree potential and the external electric field $E_{\rm{ext}}$ can be written as 
\begin{eqnarray}
    \varepsilon_{\rm{on}}^{(ij)} = \varepsilon_{\rm{H}}^{(ij)} + e E_{\rm{ext}} d_{ij}.
\end{eqnarray}
\noindent Here, we set the average of the total onsite energies $\frac{1}{N}\sum_i \varepsilon_{\rm{on}}^{(i)}$ to zero and express the total onsite energy matrix including Hartree potential and the external electric field as $H_{\rm{on}} = \text{diag}[\varepsilon_{\rm{on}}^{(1)}, \varepsilon_{\rm{on}}^{(2)}, ..., \varepsilon_{\rm{on}}^{(N)}]$ in the layer basis.

For the given top gate ($n_{\rm{tg}}$) and bottom ($n_{\rm{bg}}$) charge densities, the total carrier density $n_{\rm{tot}} = -(n_{\rm{tg}} + n_{\rm{bg}})$ is induced for the sample, and the Fermi energy $\varepsilon_{\rm{F}}$ of the sample can be calculated from $n_{\rm{tot}}$. The top-gate and bottom-gate charge densities also determine the external electric field in the sample as $E_{\rm{ext}} = \frac{2\pi e}{\epsilon_{\rm{in}}}(n_{\rm{tg}} - n_{\rm{bg}})$. By solving these self-consistently, one can obtain the onsite energy $\varepsilon_{\rm{on}}^{(i)}$.

\begin{figure}[htb]
\includegraphics[width=1.0\linewidth]{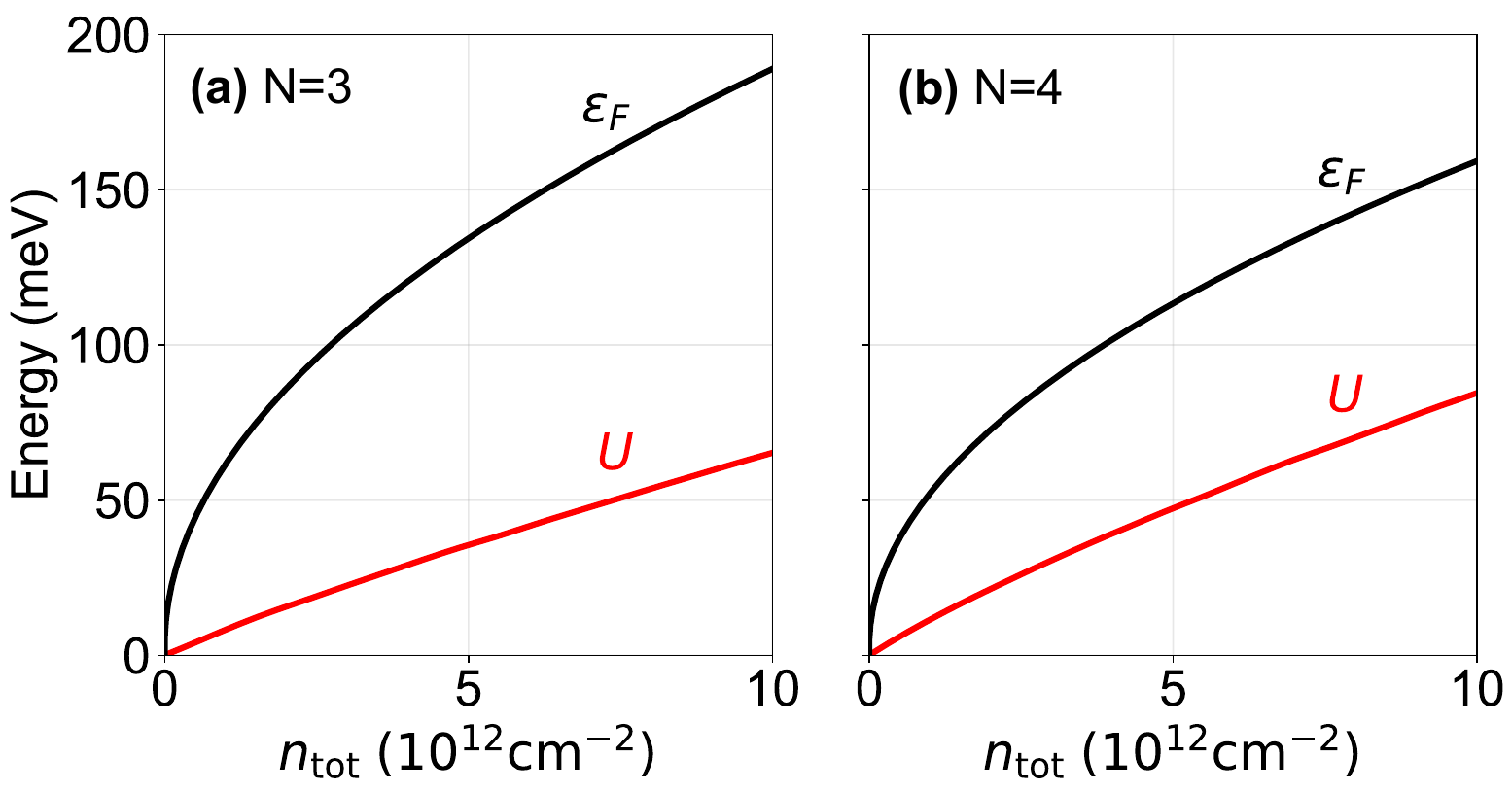}
\caption{Self-consistent Hartree calculations as a function of $n_{\rm{tot}}$ at a twist angle $\theta = 5^\circ$ in the absence of an external electric field $E_{\rm{ext}}$ for (a) $N=3$ and (b) $N=4$. The red solid line indicates the relative interlayer energy difference $U$, given by $H_{\rm{on}}=\text{diag}[-\frac{U}{3}, \frac{2U}{3}, -\frac{U}{3}]$ for $N=3$ and $H_{\rm{on}}=\text{diag}[-\frac{U}{2}, \frac{U}{2}, \frac{U}{2}, -\frac{U}{2}]$ for $N=4$, and the black solid line represents the resulting Fermi energy $\varepsilon_{\rm{F}}$.
}\label{fig: fig9}
\end{figure}

\subsection{$E_{\rm{ext}} = 0$}

In the absence of an external electric field $E_{\rm{ext}}$, the Hartree potential induced within the sample is symmetric due to the inherent symmetry of the ATMG as shown in Fig.~\ref{fig: fig9}. Specifically, this symmetry requires $\varepsilon_{\rm{H}}^{(1)}=\varepsilon_{\rm{H}}^{(3)}$ for $N=3$, and $\varepsilon_{\rm{H}}^{(1)}=\varepsilon_{\rm{H}}^{(4)}$ and $\varepsilon_{\rm{H}}^{(2)}=\varepsilon_{\rm{H}}^{(4)}$ for $N=4$. Thus, the total onsite-energy matrix in each layer with the interlayer potential energy difference $U=\varepsilon_{\rm{on}}^{(21)}$ is given by $H_{\rm{on}} = \text{diag}[-\frac{U}{3}, \frac{2U}{3}, -\frac{U}{3}]$ for $N=3$ and $H_{\rm{on}} = \text{diag}[-\frac{U}{2}, \frac{U}{2}, \frac{U}{2}, -\frac{U}{2}]$ for $N=4$. By including this effect, one can describe the resulting band structure modified by the interlayer potential difference. In Fig.~\ref{fig: fig9}, we extract $U$ for $N=3$ and $N=4$, respectively, as the total carrier density $n_{\rm{tot}}$ with $E_{\rm{ext}}=0$.

For $N=3$, we can apply perturbation theory near $\bar{K}$ with $H_{\rm{on}} = \text{diag}[-\frac{U}{3} \mathbb{I}_2, \frac{2U}{3} \mathbb{I}_6, -\frac{U}{3}\mathbb{I}_2]$ in the basis of the wave functions given in Eq.~\eqref{eq: wave functions of AT3G at K}. In this case, we obtain
\begin{eqnarray}
    H_{\bar{K}} = \begin{bmatrix}
        v_1 (\bm{k}\cdot\bm{\sigma}) - \frac{U}{3} \frac{1-24\tilde{\alpha}^2}{1+12\tilde{\alpha}^2} & 0 \\
        0 & v_2 (\bm{k}\cdot \bm{\sigma}) - \frac{U}{3}
    \end{bmatrix},
\end{eqnarray}
\noindent which implies that the two Dirac cones near $\bar{K}$ shift downward by almost the same amount, $\frac{U}{3}$ in the $\tilde{\alpha}\rightarrow 0$ regime. In contrast, at $\bar{K}'$, applying the same perturbative treatment with $H_{\rm{on}} = \text{diag}[-\frac{U}{3}\mathbb{I}_6, \frac{2U}{3}\mathbb{I}_2, -\frac{U}{3} \mathbb{I}_6]$ shifts the single Dirac cone upward by $\frac{2U}{3}$.

\begin{figure}[htb]
\includegraphics[width=1.0\linewidth]{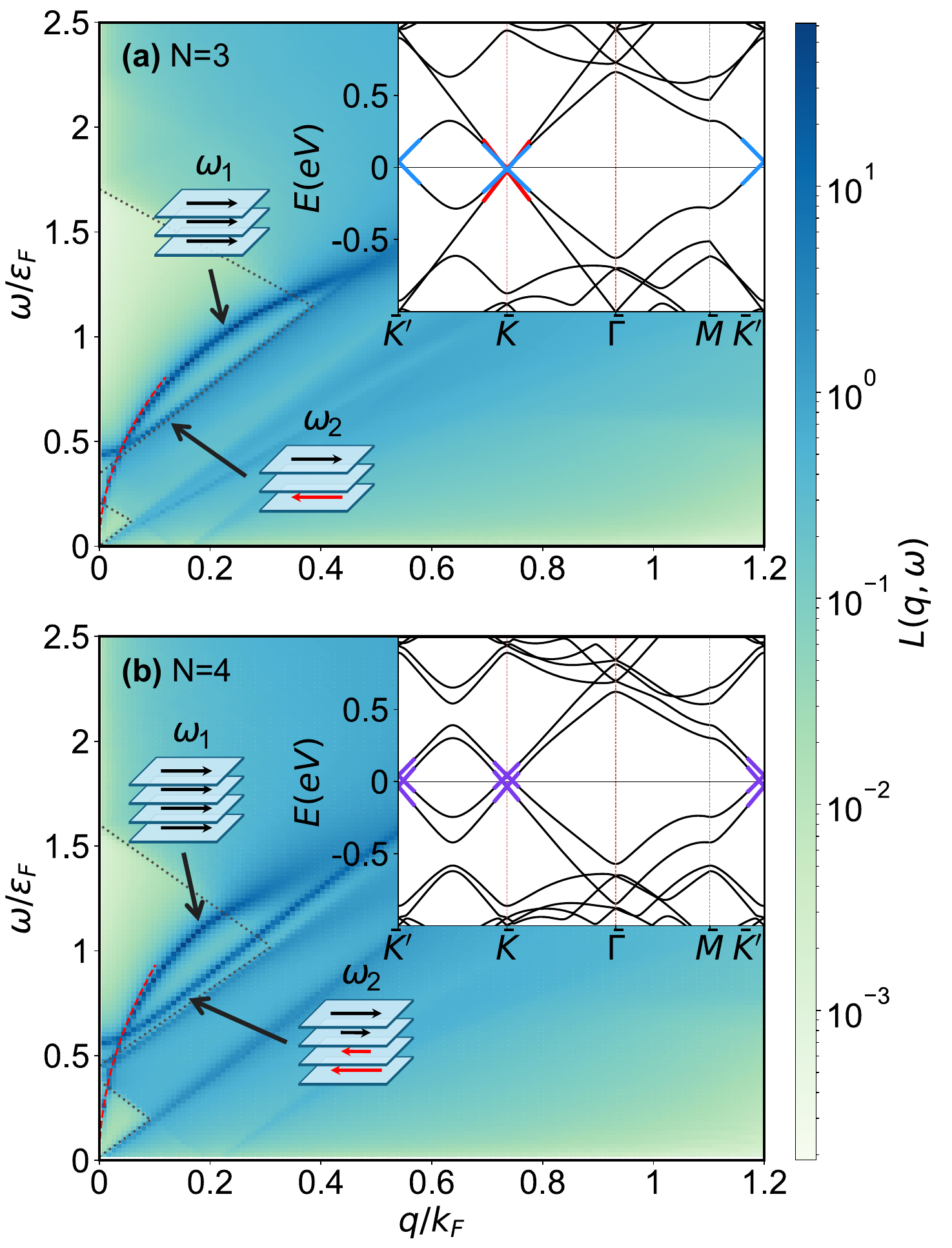}
\caption{Loss function $L(\bm{q}, \omega)$ of unbiased ATMG systems, accounting for self-consistent Hartree effects, at a twist angle $\theta = 5^{\circ}$ with $n_{\rm{tot}} = 10^{13}$ $\text{cm}^{-2}$ for (a) $N=3$ ($\varepsilon_{\rm{F}} = 0.19$ eV and $U = 0.06$ eV) and (b) $N=4$ ($\varepsilon_{\text{F}} = 0.16$ eV and $U = 0.08$ eV). The insets show corresponding band structures modified by the Hartree potentials, where the colored lines highlight the perturbative Hamiltonian near $\bar{K}$ and $\bar{K}'$.
}\label{fig: fig10}
\end{figure}

For $N=4$, evaluating the perturbation $H_{\rm{on}} = \text{diag}[-\frac{U}{2}\mathbb{I}_2, \frac{U}{2}\mathbb{I}_6, \frac{U}{2}\mathbb{I}_2, -\frac{U}{2}\mathbb{I}_6]$ near $\bar{K}$ using the basis of the wave functions provided in Eq.~\eqref{eq: wave functions of AT4G at K} yields the Hamiltonian as
\begin{eqnarray}\label{eq: perturbative Hamiltonian AT4G}
    H_{\bar{K}} = \begin{bmatrix}
        v_1 (\bm{k}\cdot\bm{\sigma}) + \frac{U}{2\sqrt{5}} & -\frac{U}{\sqrt{5}} D(\tilde{\alpha}) \\
        -\frac{U}{\sqrt{5}}D(\tilde{\alpha}) & v_2 (\bm{k}\cdot \bm{\sigma}) - \frac{U}{2\sqrt{5}}
    \end{bmatrix},
\end{eqnarray}
\noindent where
\begin{eqnarray}
    D(\tilde{\alpha}) = \frac{1-6 t_1t_2 \tilde{\alpha}^2}{\sqrt{(1+6 t_1^2\tilde{\alpha}^2)(1+6t_2^2 \tilde{\alpha}^2)}}.
\end{eqnarray} This indicates that the interlayer potential induces hybridization between the two Dirac cones near $\bar{K}$. Furthermore, applying the same perturbative analysis near $\bar{K}'$ using the wave functions from Eq.~\eqref{eq: wave functions of AT4G at K'} gives the identical Hamiltonian presented in Eq.~\eqref{eq: perturbative Hamiltonian AT4G}.

\begin{figure}[htb]
\includegraphics[width=1.0\linewidth]{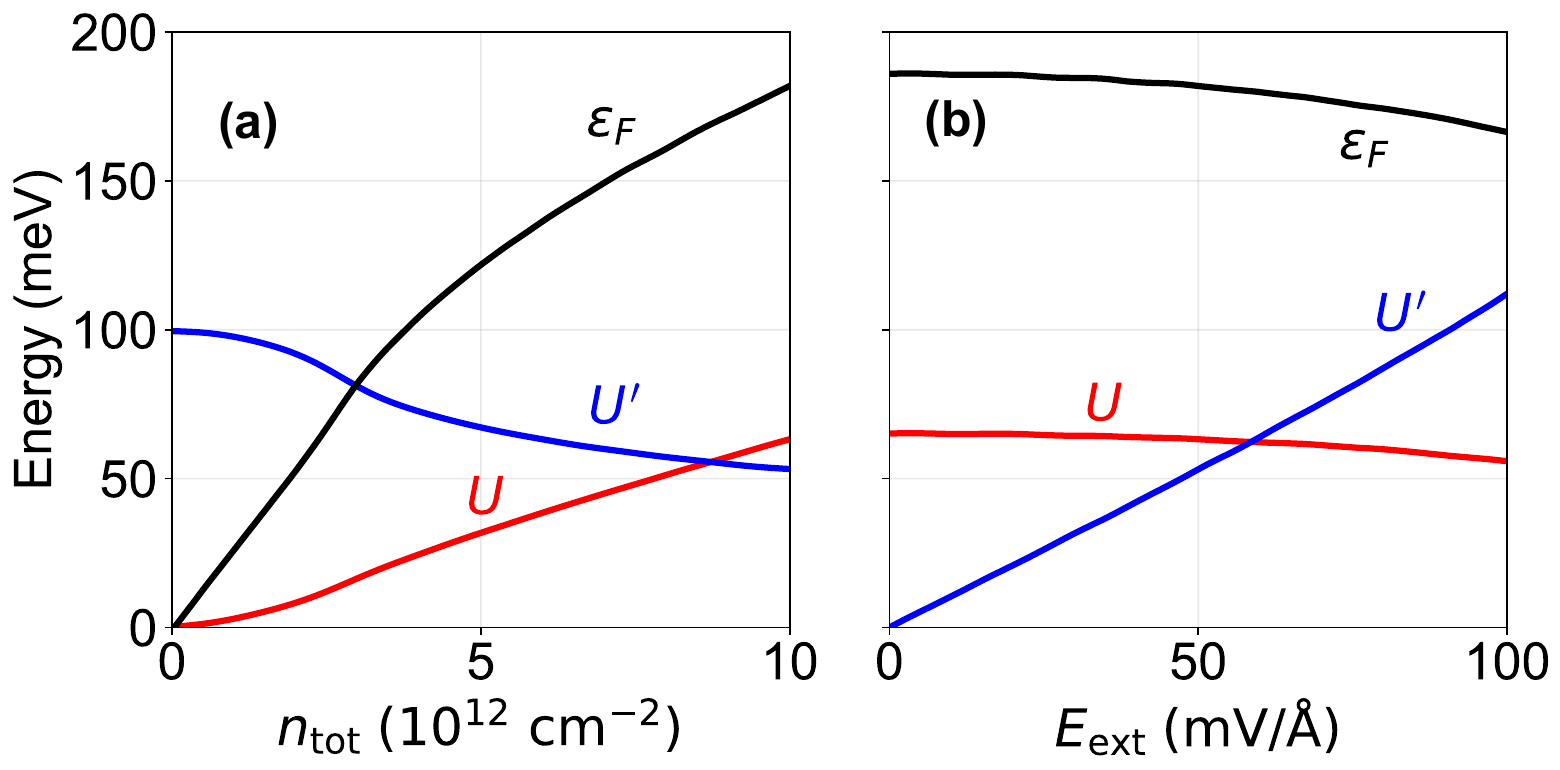}
\caption{Self-consistent Hartree calculations for $N=3$ at a twist angle $\theta = 5^\circ$ in the presence of an external electric field $E_{\rm{ext}}$. The results are plotted (a) as a function of $n_{\rm{tot}}$ at a fixed $E_{\rm{ext}} = 0.05$ V/\r{A}, and (b) as a function of $E_{\rm{ext}}$ at a fixed $n_{\rm{tot}} = 10^{13}\text{ cm}^{-2}$. The red (blue) solid line indicates the parameter $U$ ($U'$) given by $U = \frac{\varepsilon_{\rm{on}}^{(21)} +\varepsilon_{\rm{on}}^{(23)}}{2} $ ($U'=\frac{\varepsilon_{\rm{on}}^{(21)} - \varepsilon_{\rm{on}}^{(23)}}{2}$) and the black solid line represents the resulting Fermi energy $\varepsilon_{\rm{F}}$.
}\label{fig: fig11}
\end{figure}

We also compute the loss functions to account for the self-consistent Hartree effect. As illustrated in Fig.~\ref{fig: fig10}, the out-of-phase plasmon modes for $N=3$ and $N=4$ originate from interband transitions between the two Dirac cones. Consequently, the Hartree effect only alters out-of-phase modes quantitatively without changing the underlying physics. For larger $N$, however, the effect becomes more complex, and the resulting layer-dependent energy differences significantly complicate the particle-hole continuum. A complete treatment of these effects at higher $N$ could modify the plasmon damping conditions, making them density-dependent even above the critical angle.

\subsection{$E_{\rm{ext}}\neq 0$}

\begin{figure}[htb]
\includegraphics[width=1.0\linewidth]{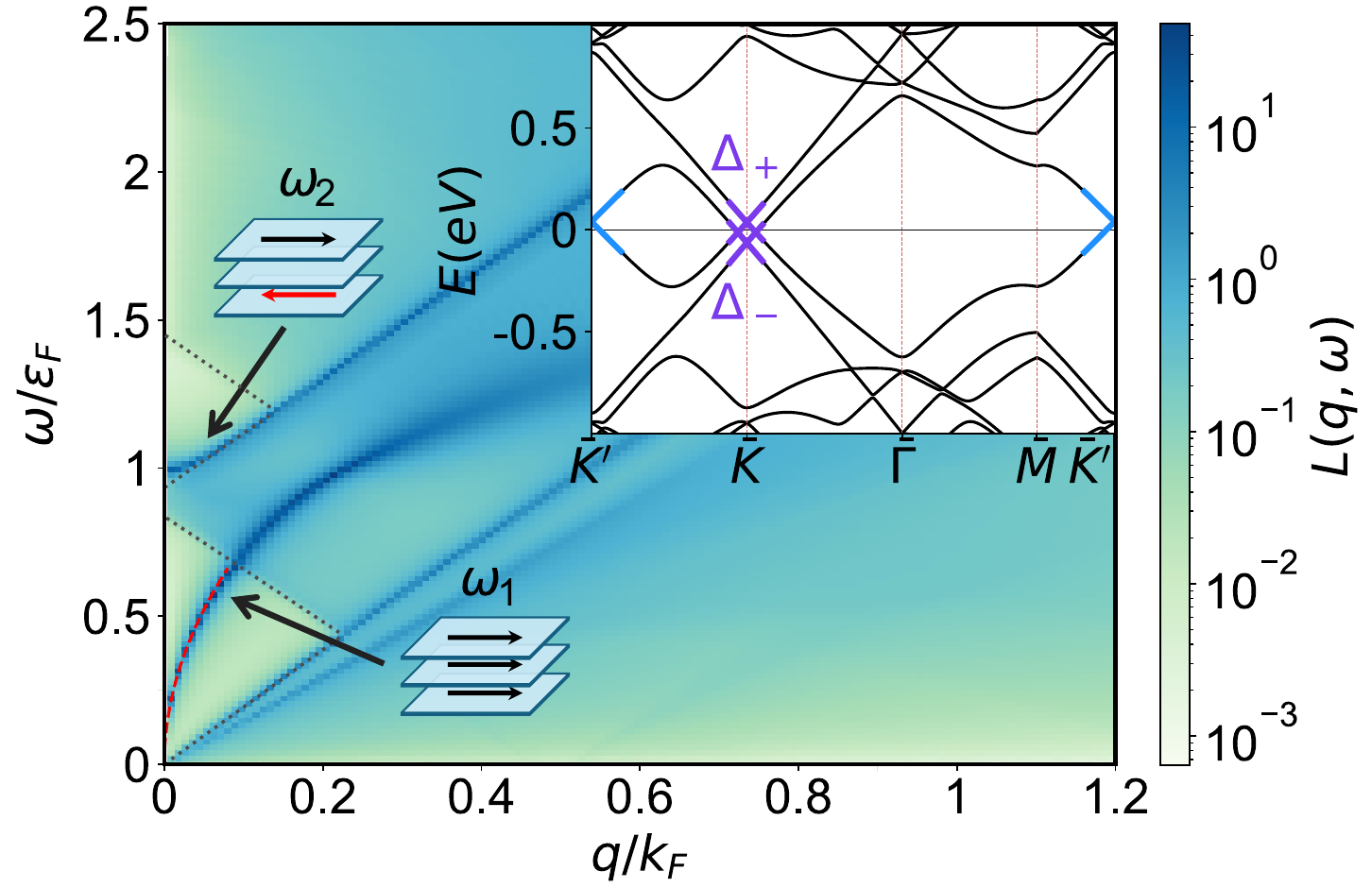}
\caption{Loss function $L(\bm{q}, \omega)$ of a biased AT3G system, accounting for self-consistent Hartree effects, at a twist angle $\theta = 5^{\circ}$ under an external electric field $E_{\rm{ext}} = 0.05$ V/Å with $n_{\rm{tot}} = 10^{13}\text{ cm}^{-2}$ ($\varepsilon_{\rm{F}} = 0.18$ eV, $U = 0.06$ eV, and $U' = 0.05$ eV).
}\label{fig: fig12}
\end{figure}

In the presence of an external electric field $E_{\rm{ext}}$, the application of the perpendicular field $(E_{\rm{ext}}\neq 0)$ induces asymmetry in the interlayer potential differences.. Due to this, the induced onsite energy $\varepsilon_{\rm{on}}^{(i)}$ can no longer be captured by a single parameter $U$. Instead, $H_{\rm{on}}$ can be described by $\varepsilon_{\rm{on}}^{(21)}$ and $\varepsilon_{\rm{on}}^{(23)}$ self-consistently. By employing perturbation theory near $\bar{K}$ within the first-shell model, we obtain
\begin{eqnarray}
    H_{\bar{K}} = \begin{bmatrix}
        v_1(\bm{k}\cdot\bm{\sigma}) - \frac{U}{3} \frac{1-24\tilde{\alpha}^2}{1+12\tilde{\alpha}^2} & -C(\tilde{\alpha}) U' \\
        -C(\tilde{\alpha})U' & v_2(\bm{k}\cdot \bm{\sigma}) - \frac{U}{3}
    \end{bmatrix},
\end{eqnarray}
\noindent where $U = \frac{\varepsilon_{\rm{on}}^{(21)} +\varepsilon_{\rm{on}}^{(23)}}{2} $ and $U'=\frac{\varepsilon_{\rm{on}}^{(21)} - \varepsilon_{\rm{on}}^{(23)}}{2}$. In Fig.~\ref{fig: fig11}, we extracted $U$ and $U'$ self-consistently. Notice that this Hamiltonian is identical to Eq.~\eqref{eq: perturbation theory AT3G}, as the hybridization of the two Dirac cones results in an AA stacking band structure due to $U'$ that is overall shifted by $-\frac{U}{3}$ in the $\tilde{\alpha}\rightarrow 0$ regime. Furthermore, an analogous perturbative treatment at $\bar{K}'$ shifts the single Dirac cone upward by $\frac{2}{3}U$. As illustrated in Fig.~\ref{fig: fig12}, these findings demonstrate that our analysis in the main text remains valid even under realistic dual-gating conditions, confirming that the gapped out-of-phase plasmon is generated by the interband transitions between the two Dirac cones and can be tuned via a perpendicular electric field.

\bibliography{references}

\end{document}